\documentclass{raa}           
\usepackage{graphicx,times}
\usepackage{natbib}
\usepackage{multirow}
\usepackage{amssymb,amsmath}

\bibpunct{(}{)}{;}{a}{}{,}
\usepackage[a4paper=true,pagebackref=true]{hyperref}
\hypersetup{colorlinks = true, linkcolor = blue, anchorcolor = red, citecolor = blue, filecolor = red, pagecolor = red, urlcolor = red}

\begin{document}

\title{Are there magnetars in high-mass X-ray binaries?$^*$
\footnotetext{\small $*$ Supported by the National Natural Science Foundation of China.}
}

\volnopage{ {\bf 20XX} Vol.\ {\bf X} No. {\bf XX}, 000--000}
\setcounter{page}{1}

\author{Kun Xu\inst{1,2,3}, Xiang-Dong Li\inst{3,4}, Zhe Cui\inst{3}, Qiao-Chu Li\inst{3,4}, Yong Shao\inst{3,4}, Xilong Liang\inst{1,2}, Jifeng Liu\inst{1,2,5}}

\institute{ School of Astronomy and Space Sciences, University of Chinese Academy of Sciences, Beijing, People’s Republic of China; {\it xukun@smail.nju.edu.cn} \\
	\and 
	Key Laboratory of Optical Astronomy, National Astronomical Observatories, Chinese Academy of Sciences, Beijing, People’s Republic of China \\
	\and 
	Department of Astronomy, Nanjing University, Nanjing 210023, People’s Republic of China \\
	\and 
	Key Laboratory of Modern Astronomy and Astrophysics, Nanjing University,
	Ministry of Education, Nanjing 210023, People’s Republic of China \\
	\and 
	WHU-NAOC Joint Center for Astronomy, Wuhan University, Wuhan, People’s Republic of China \\
	\vs \no
	{\small Received 20XX Month Day; accepted 20XX Month Day}
}

\abstract{Magnetars form a special population of neutron stars with strong magnetic fields and long spin periods. 
About 30 magnetars and magnetar candidates known currently are probably isolated. 
But the possibility that magnetars are in binaries hasn't been excluded. 
In this work, we perform spin evolution of neutron stars with different magnetic fields in wind-fed high-mass X-ray binaries and compare the spin period distribution with observations, aiming to find magnetars in binaries.
Our simulation shows that some of the neutron stars, which have long spin periods or in wide-separation systems, need strong magnetic fields to explain their spin evolution.
This implies that there are probably magnetars in high-mass X-ray binaries.
Moreover, this can further provide a theoretical basis for some unclear astronomical phenomena, such as the possible origin of periodic fast radio bursts from magnetars in binary systems.
\keywords{stars: magnetars  --- stars: neutron --- stars: rotation --- stars: supergiants --- accretion, accretion discs --- X-rays: binaries}
}

\authorrunning{K. Xu et al. }            
\titlerunning{Are there magnetars in high-mass X-ray binaries?}  
\maketitle

%
\section{Introduction}           
\label{sect:intro}

Magnetars are thought to be a special population of neutron stars (NSs) with magnetic fields $B \gtrsim 10^{14}$ G \citep{Duncan1992,Kaspi2017} in general.
They pulsate in X-ray band as persistent emission sources with bursts and outbursts, which are powered by their strong magnetic fields \citep{An2019,Beniamini2019,Kaspi2017}.
While low-field magnetars were found in observations \citep{Rea2010,Zhou2014}, 
and the dipole magnetic field of a few $\times 10^{12}$ G indicates that they are possibly  old magnetars \citep{Turolla2011,DallOsso2012,Tong2012}.
About 30 magnetars and candidates known currently are listed in the McGill magnetar catalog \footnote{The url of the catalog is http://www.physics.mcgill.ca/$\sim$pulsar/magnetar/main.html} \citep{Olausen2014} with  characteristic magnetic fields from $\sim 6 \times 10^{12}$ G to $\sim 2 \times 10^{15}$ G. 
In addition, some central compact objects (CCOs) may be magnetar candidates, e.g., the one located close to the center of RCW 103 (1E 161348$-$5055) is thought to be a magnetar with $B \gtrsim 5 \times 10^{15}$ G \citep{DeLuca2006,Ho2017,Tong2016,Xu2019}.
All magnetars and candidates are probably isolated NSs because no optical/near-infrared counterpart of them has been identified \citep{King2019}.
But in the theory of binary evolution, one can't exclude the possibility that magnetars are in binaries.

So here comes the question, are there magnetars in binary systems? 
It seems to be answered when the first ultraluminous X-ray pulsar (ULXP) M82 X-2 was found \citep{Bachetti2014}.
The peaked luminosity of M82 X-2 is $\sim 10^{40}$ erg s$^{-1}$, so its magnetic field has been estimated to be $\sim 10^{14}$ G \citep{Eksi2015,Tsygankov2016}.
While other estimation implies that its magnetic field is $\sim 10^{12}-10^{13}$ G \citep{DallOsso2015,Xu2017} 
or even lower ($\lesssim 10^9$ G, \cite{Kluzniak2015}).
Then more and more ULXPs are discovered but none of them was confirmed to be a magnetar \citep{King2019}.
However, an interesting idea is that M82 X-2 is an accreting low magnetic field magnetar 
\citep{Tong2015,Tong2019} with $B \sim 10^{12}$ G \citep{Chen2017}.
This is in accord with the current view that the magnetars' ultrahigh fields decay on a timescale $< 10^5-10^6$ yr by Hall draft and Ohm diffusion \citep{Turolla2015}. 
After decay, the magnetic fields of old magnetars (or low magnetic field magnetars, or post-magnetars) are in the similar magnitude as that of other pulsars'. Therefore, the problem now is how to pick old magnetars out from the zone of NSs.
The long period ($\sim 2.6$ h) X-ray pulsar in the high-mass X-ray binary (HMXB) 4U 0114+65 provides some inspirations, which could be an old magnetar \citep{Li1999,Sanjurjo-Ferrrin2017} with age of 2.4$-$5 Myr \citep{Igoshev2018}. 
\cite{Li1999} thought that it was born as a magnetar and spun down to $\sim 10$ s by magnetic dipole radiation on a timescale of $10^4-10^5$ yr, then its magnetosphere started to interact with the wind material from the companion and the NS spun down to $\sim 10^4$ s within $10^5$ yr.
This can lead one to find some clues to the problem in wind-fed HMXBs.

Some mysterious electromagnetic radiations can be interpreted by magnetars in binaries, such as fast radio bursts (FRBs).
FRBs are millisecond-duration and extremely bright radio transients \citep[for reviews, see][]{Cordes2019, Petroff2019, Zhangbing2020, Xiao2021}.
Since 2007, when the first FRB was discovered, there have been a large number of models proposed \citep[e.g.,][]{Lyubarsky2014, Geng2015, Dai2016, Zhangbing2017, Yang2018, Yang2021, Platts2019}. However, the radiation mechanism of FRBs is still unclear.
Remarkably, FRB 200428 was detected in association with an X-ray burst from the Galactic magnetar SGR 1935+2146 \citep[e.g.,][]{Bochenek2020, CHIME2020b, Mereghetti2020, LiCK2021, Tavani2021}. This indicate that magnetars can be a striking scenario for the origin of FRBs \citep[e.g.,][]{Zhangbing2020}.
More interestingly, there are some repeating FRBs showing long periodic activities, e.g. $\sim$ 16.35 days for FRB 180916.J0158+65 \citep{CHIME2020a} and a possible $\sim160$-day for FRB 121102 \citep{Rajwade2020, Cruces2021}. 
Many scenarios have been proposed to explain the periodicities \citep[e.g.,][]{Smallwood2019, Beniamini2020, Dai2020, Gu2020, Levin2020, Lyutikov2020, DengCM2021, GengJJ2021, Kuerban2021, LiDZ2021, Wada2021, XuK2021}, in which a leading one invokes magnetars in high-mass binaries \citep[e.g.,][]{Ioka2020, LiQC2021}. 
Therefore, it also becomes a very interesting subject in the FRB field whether magnetars can exist in binary systems \citep{ZhangXF2020}.

\cite{Popov2012} have modeled the formation channel for the NS in the Be/X-ray binary SXP 1062, which has a long spin period ($P_{\rm s}=1062$ s) and short age, indicating that its initial magnetic field may be larger than $10^{14}$ G and decayed to $10^{13}$ G.
\cite{Zhang2004} have calculated the spin evolution of NSs in OB/X-ray binaries before steady wind accretion.
\cite{Shakura2012} proposed that wind matter around an accreting NS should form a quasi-spherical shell above the magnetosphere rather than directly accreted if the cooling time of the wind plasma is longer than its free-fall time.
The results of population synthesis \citep{Li2016} with the subsonic settling accretion \citep{Shakura2012} are better consistent with observations.
\cite{Karino2020} found that the wind velocity makes a great difference on the spin evolution of NSs in wind-fed HMXBs.

In this paper,
we perform the spin evolution of neutron stars in wind-fed high-mass X-ray binaries and 
compare the spin periods distribution with the observations, aiming to see if there is any difference between old magnetars and other NSs.
In Section 2, we describe the wind accretion regimes and Section 3 presents the numerical results as well as the comparison with observations.
Finally, our discussion and summary are shown in Section 4 and 5, respectively.

\section{Model}

\subsection{Wind Accretion Regimes}\label{wind_acc_reg}

\begin{figure} 
	\centering
	\includegraphics[width=14.0cm, angle=0]{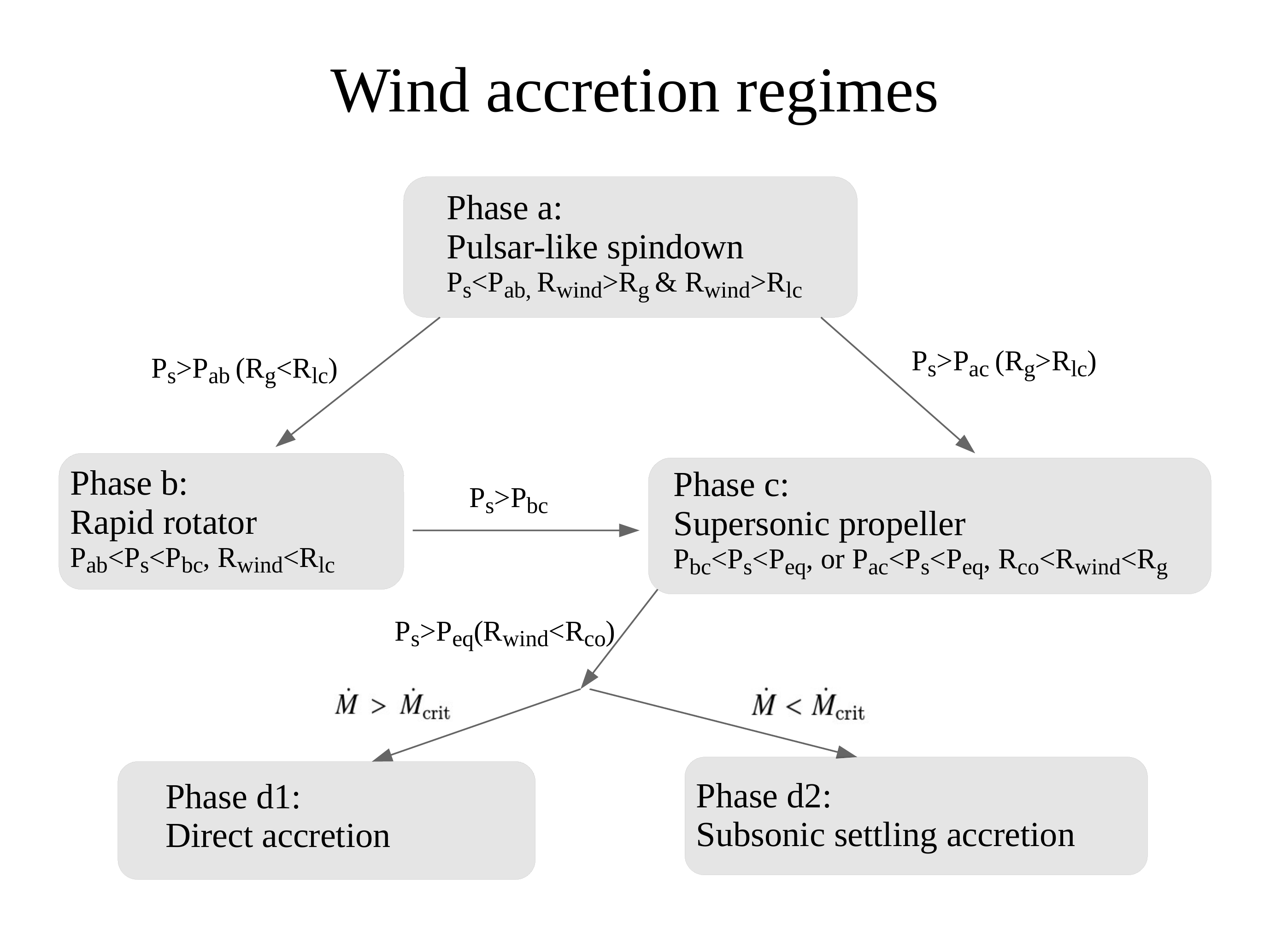}
	\caption{Wind accretion regimes, to see the section \ref{wind_acc_reg}. } 
	\label{fig:regime}
\end{figure}

We consider a wind-fed accretion binary system consisting of a NS of mass $M_{\rm NS}=1.4 M_{\odot}$ and a massive main-sequence (MS) companion without filling its Roche lobe (RL).
In general, the NS in a binary system is born with a short spin period ($P_{\rm s} \sim 0.01-0.1$ s), and then it can experience the following possible phases sketched in Figure \ref{fig:regime}.

\subsubsection{Phase a: the Pulsar-like Spindown Phase}

First, an NS appears as a radio pulsar after its birth in supernova explosion if its magnetic or radiation pressure is high enough to boot the wind plasma outside the Bondi radius $R_{\rm g} = 2GM_{\rm NS}/v_{\rm rel}^2$
\citep{Bondi1944} or the radius of the light cylinder $R_{\rm lc} = c P_{\rm s}/2\pi$.
Here $v_{\rm rel}=(v_{\rm orb}^2+v_{\rm w}^2)^{1/2}$ is the relative speed towards the NS, where $v_{\rm orb}$ is the NS velocity at the orbit and $v_{\rm w}$ is the stellar wind velocity.
$G$ and $c$ are the gravitational constant and the speed of light, respectively.
For clarity, we define $R_{\rm wind}$ as the wind termination radius from the neutron star and $R_{\rm eff}=max(R_{\rm g}, R_{\rm lc})$ as the effective radius of the NS, 
so can get $R_{\rm wind}>R_{\rm eff}$ in phase a.
The spin-down rate induced by magnetic dipole radiation in this phase is \citep{Zhang2004}
\begin{equation}
	\dot{P}_{\rm a} = 9.73 \times 10^{-18} \mu_{31}^{2} P_{\rm 0}^{-1} I_{45}^{-1} {\rm \ s \ s^{-1}},
\end{equation}
where $\mu_{31}$ and $P_{\rm 0}$ are the NS's magnetic dipole moment in units of $10^{31}$ G cm$^2$ and its spin period in units of 1 s, respectively.
$I = I_{45} \times 10^{45} {\rm ~g ~cm}^2$ is the moment of inertia of the NS.

Phase a ends when either the wind material penetrates the light cylinder radius or it enters the Bondi radius, 
i.e., $R_{\rm wind}<R_{\rm eff}$, 
correspondingly the transitional spin periods $P_{\rm ab}$ and $P_{\rm ac}$ \citep{Davies1981}
\begin{equation}
	P_{\rm ab} = 0.37 \mu_{31}^{1/3} \dot{M}_{15}^{-1/6} v_8^{-5/6} \left(\frac{M_{\rm NS}}{M_\odot}\right)^{1/3} {\rm \ s},
\end{equation}
and 
\begin{equation}
	P_{\rm ac} = 0.38 \mu_{31}^{1/2} \dot{M}_{15}^{-1/4} v_8^{-1/2} {\rm \ s},
\end{equation}
where $v_8$ is the relative speed towards the NS in units of $10^8$ cm s$^{-1}$ and $\dot{M}=\dot{M}_{15} \times 10^{15}$ g s$^{-1}$
is the accretion rate onto the NS given by \cite{Bondi1944}
\begin{equation}
	\dot{M} = \pi R_{\rm g}^2 \rho_{\rm w} v_{\rm rel}.
\end{equation}
Here $\rho_{\rm w}$ is the stellar wind density at the orbit of the NS, which can be computed by assuming an isotropically expanding wind at a speed of $v_{\rm w}$
\begin{equation}
	\rho_{\rm w} = -\dot{M}_2 / (4\pi a^2 v_{\rm w}),
\end{equation}
where $a$ is the orbital separation of the system and $\dot{M}_2$ is the mass loss rate of the donor star.

\subsubsection{Phase b: the Rapid Rotator Phase}

The rapid rotator phase (phase b) comes if the wind material from the donor star enters the light cylinder of the NS in front of the Bondi radius in the weak stellar wind case \citep{Illarionov1975},
i.e., $R_{\rm g}<R_{\rm lc}$ in this case and phase b occurs when $R_{\rm wind}<R_{\rm lc}$, as well as $P > P_{\rm ab}$ commences before $P > P_{\rm ac}$ if $P_{\rm ab} < P_{\rm ac}$.
The spin-down rate in phase b is \citep{Zhang2004} 
\begin{equation}
	\dot{P}_{\rm b} = 8.59 \times 10^{-17} \mu_{31} \dot{M}_{15}^{1/2} v_8^{5/2} P_{\rm 0}^{2} {\rm \ s \ s^{-1}}.
\end{equation}

When the outer boundary of the wind plasma reaches the Bondi radius ( i.e., $R_{\rm wind}=R_{\rm g}$), phase b stops at \citep{Davies1981}
\begin{equation}
	P_{\rm bc} = 0.22 \mu_{31} \dot{M}_{15}^{-1/2} v_8^{1/2} \left(\frac{M_{\rm NS}}{M_\odot}\right)^{-1} {\rm \ s}.
\end{equation}

\subsubsection{Phase c: the Supersonic Propeller Phase}

When $R_{\rm wind}<R_{\rm g}$, which is equivalent to $P > P_{\rm bc}$, the supersonic propeller phase (phase c) arrives in the weak stellar wind case.
While neutron stars will enter phase c directly from phase a if the wind stellar is strong enough \citep{Illarionov1975}, i.e., 
phase c comes when $R_{\rm wind}<R_{\rm g}$ in the case of $R_{\rm g}>R_{\rm lc}$ as well as $P > P_{\rm ac}$ occurs before $P > P_{\rm ab}$ if $P_{\rm ac} < P_{\rm ab}$.
The spin-down rate in this phase is \citep{Zhang2004}
\begin{equation*}
	\dot{P}_{\rm c} = 1.6 \times 10^{-14} \mu_{31}^{8/13} \dot{M}_{15}^{9/13}  P_{\rm 0}^{21/13} {\rm \ s \ s^{-1}}.
\end{equation*}

When the inner boundary of the wind plasma reaches the corotation radius $R_{\rm co} = (GM/\Omega_{\rm s}^2)^{1/3}$, phase c ends at the equilibrium spin period \citep{Davies1981}
\begin{equation}
	P_{\rm eq} = 5.0 \mu_{31}^{2/3} \dot{M}_{15}^{-1/3} v_8^{-2/3} {\rm \ s}.
\end{equation}

\subsubsection{Phase d1 and Phase d2: the Accretion Phases}

When the inner boundary of the wind material becomes smaller than the the corotation radius ($R_{\rm wind}<R_{\rm co}$) and the cooling is efficient, indicated by $P_{\rm s} > P_{\rm eq}$, a bow shock can form at the Bondi radius around the NS and quasi-spherical accretion onto the NS occurs.
The shocked matter falls towards the magnetosphere of the NS at a supersonic speed if it cools down rapidly, or it forms a quasi-static shell around the magnetosphere and settles down subsonically if the matter remains hot,
corresponding to the Bondi-Hoyle-Littleton (BHL) accretion phase (phase d1) or the subsonic settling accretion phase (phase d2).
Phase d1 takes place at $\dot{M}>\dot{M}_{\rm crit}$ while phase d2 occurs at $\dot{M}<\dot{M}_{\rm crit}$\citep{Shakura2012}, where $\dot{M}_{\rm crit} = 8.4 \times 10^{15} \mu_{31}^{1/4} (M/M_\odot)^{-1/2} R_6^{7/8} {\rm ~g ~s}^{-1}$ \citep{Arons1976a,Arons1976b,Elsner1977,Elsner1984}.

In phase d1, the accreted material transfers spin-up torque to the NS
\begin{equation}
	N_{\rm super}=I\dot{\Omega} = \dot{M} R_{\rm M}^2 \Omega_{\rm S},
\end{equation}
where $R_{\rm M}=[\mu^4/(2GM_{\rm NS}\dot{M}^2)]^{1/7}$ is the magnetospheric radius.
This phase ends at the spin equilibrium state with $P_{\rm s}=P_{\rm eq}$.
In phase d2, besides the material torque, the interaction between the magnetosphere and the plasma exerts an additional one on the NS.
Therefore the total torque that governs the NS spin evolution is 
\begin{equation}
	N_{\rm sub} = A_{\rm const} \dot{M}_{\rm X, 16}^{7/11} - B_{\rm const} \dot{M}_{\rm X, 16}^{3/11},
\end{equation}
where 
\begin{equation}
	A_{\rm const} \simeq 3.73 \times 10^{31} K_1 \mu_{31}^{1/11} v_8^{-4} \left(\frac{P_{\rm orb}}{10 {\rm \ day}}\right)^{-1},
\end{equation}
\begin{equation}
	B_{\rm const} \simeq 3.61 \times 10^{31} K_1 \mu_{31}^{13/11} \left(\frac{P_{\rm s}}{100 {\rm \ s}}\right)^{-1}.
\end{equation}
Here $K_1 \sim 40$ is a dimensionless numerical factor \citep{Shakura2012} and $\dot{M}_{\rm X, 16}=\dot{M}_{\rm X} / 10^{16} {\rm ~g ~s}^{-1}$, where $\dot{M}_{\rm X}=10^7 {\rm ~g~s}^{-1} \cdot \mu_{30}^{2/21} (\dot{M}/10^{10} {\rm ~g~s}^{-1})^{9/7}$ \citep{Popov2015}.
The equilibrium spin period in phase d2, where this phase stops, reads \citep{Li2016}
\begin{equation}
	P_{\rm eq,sub} \simeq 96.8 \mu_{31}^{12/11} v_8^4 \dot{M}_{16}^{-4/11} \left(\frac{P_{\rm orb}}{10 {\rm \ day}}\right) {\rm \ s}.
\end{equation}
In observation, the system in accretion phases (phase d1 and phase d2) can be seen in X-ray band if the luminosity is high enough.

\subsection{The Evolution of the Binary}

We use the binary star evolution (BSE) code \citep{Hurley2000,Hurley2002,Kiel2006,Shao2014,Shao2015,Shao2021} to simulate the evolution of the binary systems.
The simulation begins from the birth time of the NS when the optical star is thought to be in the zero-main-sequence stage, 
and ends when the optical/companion star starts to fill its Roche lobe or explodes as a supernova.

One can estimate the wind velocity adopting the standard formula from \cite{Castor1975}
\begin{equation}
	v_{\rm w} = \eta v_{\rm esc} (1-R_2/a)^{\beta},
\end{equation}
where $v_{\rm esc}=\sqrt{2GM_2/R_2}$ is the escape velocity at the surface of the optical star.
Here we take the coefficient $\eta \sim 0.5-3$ and the power law index $\beta \sim 0.8 - 7$ \citep{Waters1989,Owocki2014,Karino2020}.

\subsection{The Evolution of the Magnetic Field}

\begin{figure} 
	\centering
	\includegraphics[width=14.0cm, angle=0]{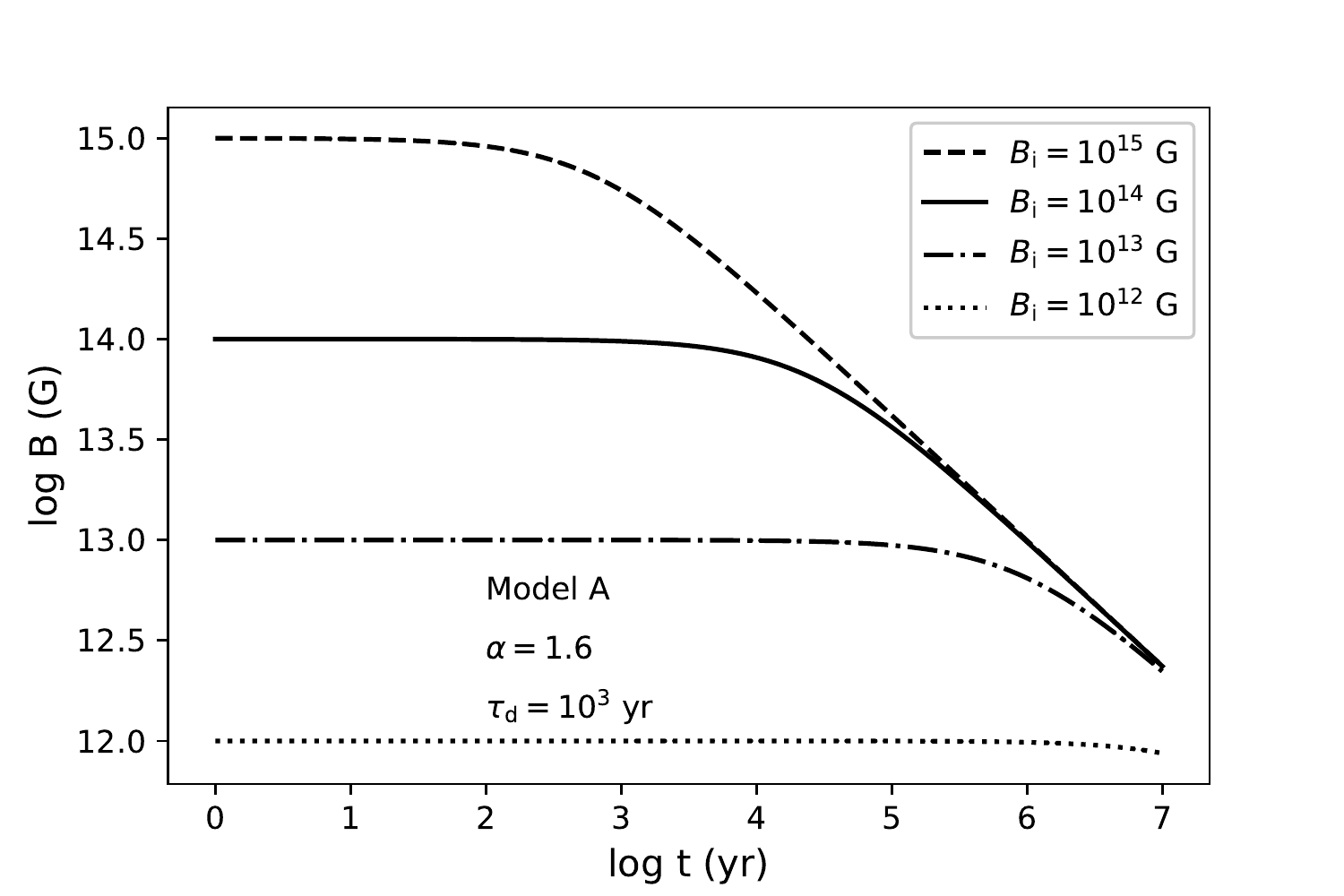}
	\caption{The magnetic field evolution with $B_{\rm i}=10^{12}$, $B_{\rm i}=10^{13}$, $B_{\rm i}=10^{14}$ and $B_{\rm i}=10^{15}$ G, where $B< 10^{13}$ G remains unchanged in its life
	while $B_{\rm i} \geq 10^{13}$ G decays in the form of Eq\eqref{eq:B_evo}. } 
	\label{fig:B_evo}
\end{figure}

The magnetic field of an NS decays in the following form \citep{Aguilera2008,DallOsso2012,Fu2012}
\begin{equation}\label{eq:B_evo}
	B(t)=B_{\rm i} (1+ \alpha t/\tau_{\rm d,i})^{-1/\alpha},
\end{equation}
where $\alpha=1.6$ and the initial field decay time $\tau_{\rm d,i}=10^{3} {\rm ~yr}/(B_{\rm i}/10^{15} {\rm ~G})^{\alpha}$ \citep{Fu2012} in our reference model.
We plot the the magnetic fields evolution with four initial values in Figure \ref{fig:B_evo}, which are $B_{\rm i}=10^{12}$, $B_{\rm i}=10^{13}$, $B_{\rm i}=10^{14}$ and $B_{\rm i}=10^{15}$ G.

\section{Results}

\subsection{The Reference Model}

\begin{figure}
	\centering
	\includegraphics[width=14.0cm, angle=0]{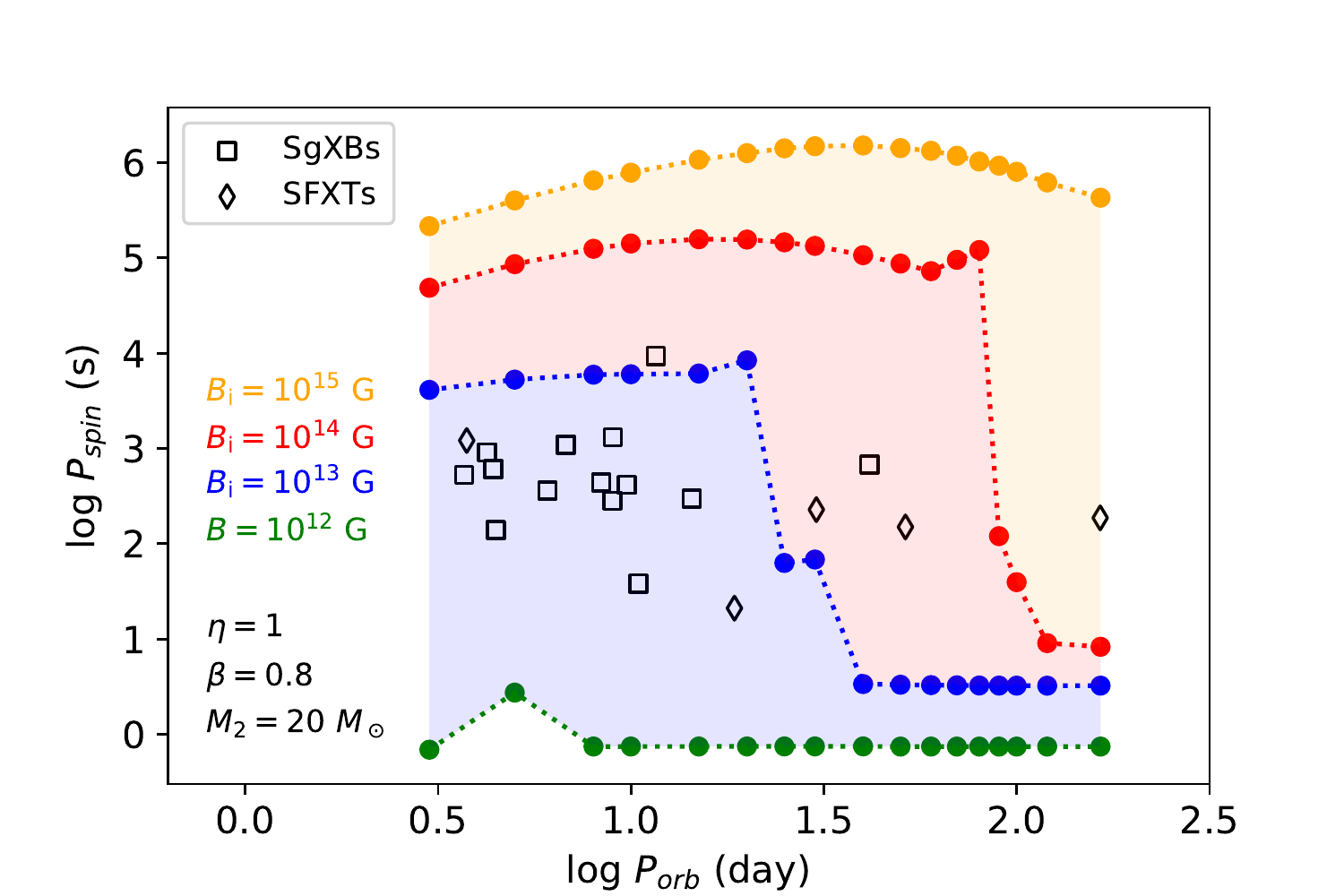}
	\caption{Results of the reference model in the Corbet diagram. 
	The filled circles indicate the maximum spin periods $P_{\rm s,max}$ during the evolution process in each case.
	The green, blue, red and orange circles represent that the initial magnetic fields of the NSs are $B_{\rm i}=10^{12}$, $B_{\rm i}=10^{13}$, $B_{\rm i}=10^{14}$ and $B_{\rm i}=10^{15}$ G, respectively.
	The dashed lines string the circles in the same colour and the coloured regions give the predicted $P_{\rm s,max}$ distributions for the NSs with initial magnetic field between the upper and lower dashed lines. 
	The squares and diamonds mark two subclasses of HMXBs, which are SgXBs and SFXTs.}
	\label{fig:m20_bt}
\end{figure}

\begin{figure}
	\includegraphics[width=0.5\textwidth]{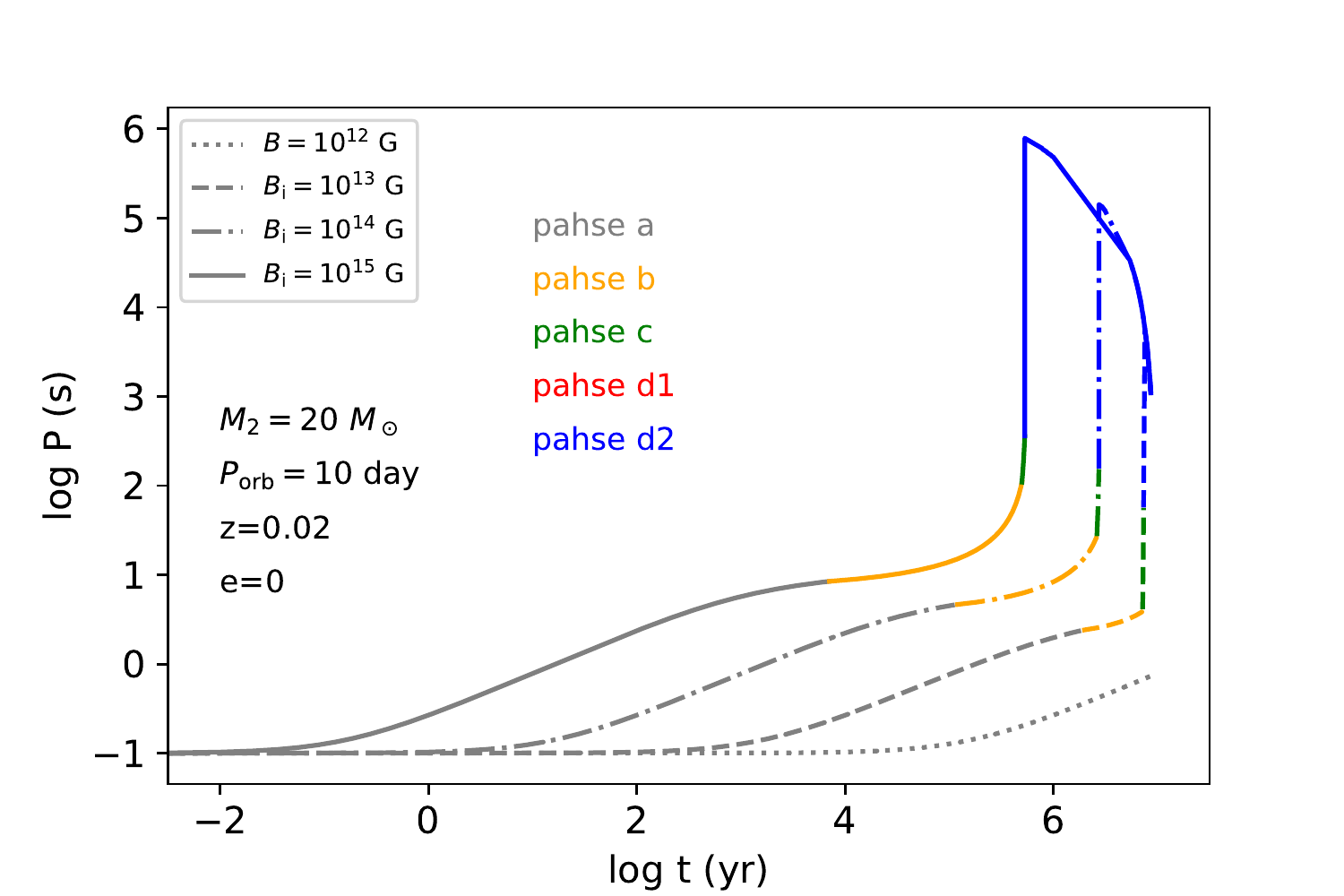}
	\includegraphics[width=0.5\textwidth]{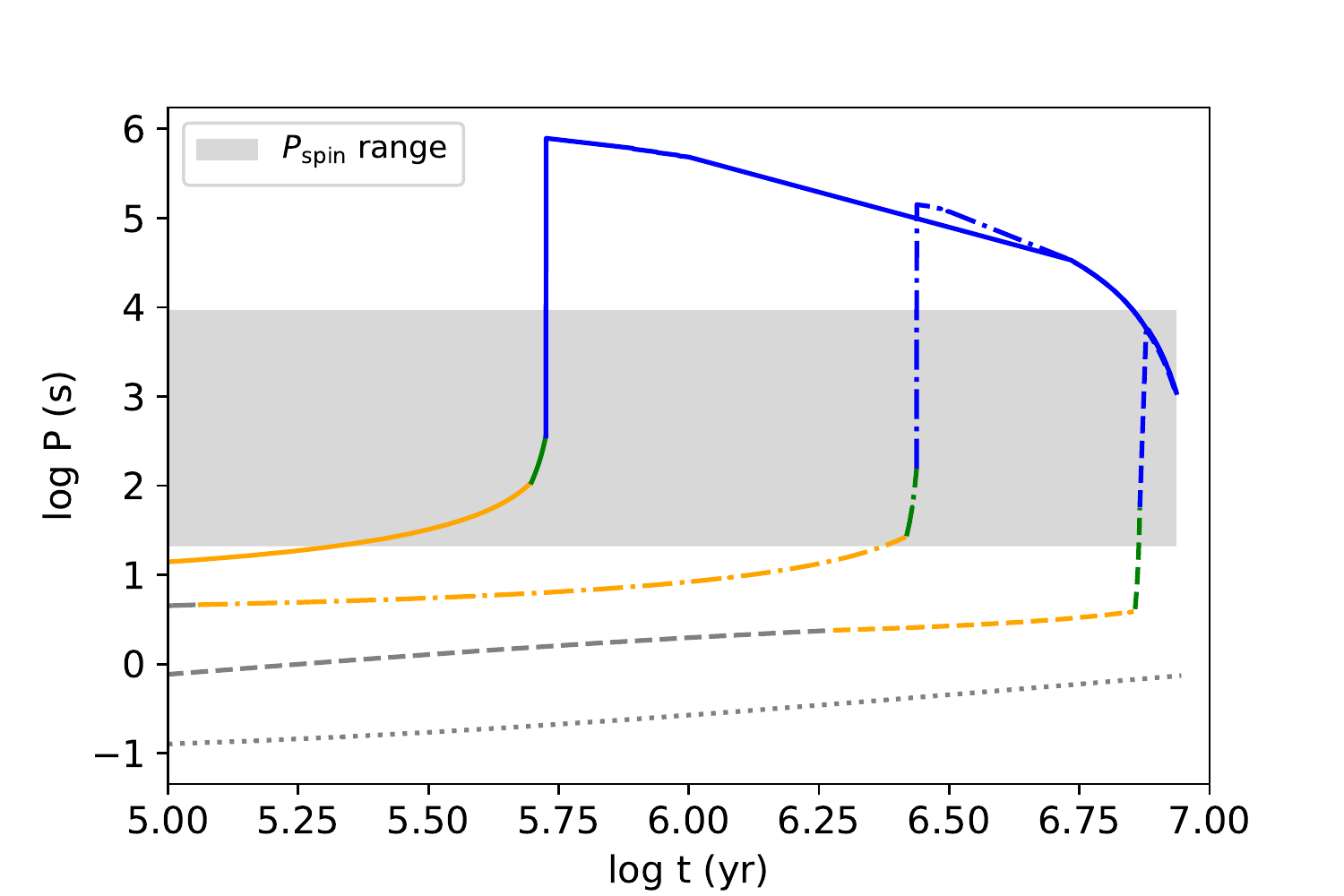}
	\caption{The spin evolution process of the reference model when $P_{\rm orb} = 10$ day. 
	The dotted, dashed , dot-dashed and solid lines represent the evolution of NSs with magnetic field $B_{\rm i}=10^{12}$, $B_{\rm i}=10^{13}$, $B_{\rm i}=10^{14}$ and $B_{\rm i}=10^{15}$ G, respectively. 
	In each line, the black, orange, green, red and blue parts indicate that the NSs are in phase a, b, c, d1 and d2.
	In phase c (the supersonic propeller phase) and d2 (the subsonic settling accreion phase) before it reaches equilibrium, the NS loses angular momentum rapidly so that its spin period rises steeply.
	The right panel is the magnified part of the left panel with $5<{\rm log} t ~ ({\rm yr})<7$ and the gray region covers the $P_{\rm spin}$ range of the NSs in HMXBs. }
	\label{fig:m20_po10}
\end{figure}

We first consider a series of circular orbital binaries as the reference model, in which each one consists of a $1.4 {\rm ~} M_\odot$ neutron star and 
a $20 {\rm ~} M_\odot$ optical star with metallicity $Z=Z_\odot$.
The wind parameters of the companion $\eta$ and $\beta$ are set up to be $1.0$ and $0.8$ respectively.
Figure \ref{fig:m20_bt} shows the results of the reference model in the Corbet diagram, where the filled circles indicate the maximum spin periods $P_{\rm s,max}$ during the evolution process in each case.
From lower to upper, the green, blue, red and orange circles represent that the initial magnetic fields of the NSs are $B_{\rm i}=10^{12}$, $B_{\rm i}=10^{13}$, $B_{\rm i}=10^{14}$ and $B_{\rm i}=10^{15}$ G, respectively.
From left to right in each colour, the circles symbolize the orbital periods of the binaries, which are 3, 5, 8, 10, 15, 20, 25, 30, 40, 50, 60, 70, 80, 90, 100, 120, 165 day, respectively.
The dashed lines string the circles in the same colour,
predicting the possible $P_{\rm s,max}$ with the $P_{\rm orb}$ ranging from $3$ to $165$ day.
The coloured regions give the predicted $P_{\rm s,max}$ distributions of the NSs with initial magnetic field between the upper and lower dashed lines. 
For example, the blue dashed line and region predict the $P_{\rm s,max}$ of the NSs with $B_{\rm i}=10^{13}$ G and $10^{12} < B_{\rm i} \leq 10^{13}$ G.
From this figure one can know that the NSs with high initial magnetic field and in short orbital period systems can spin down to very slow rotation, because the strong magnetic fields and the small separations make the NS easy to capture stellar wind from the companion.
The NSs soon enter phase c or d2 after captured and quickly lose their angular momentum (to see Figure \ref{fig:m20_po10}).
There are significant drop in the blue and red dashed lines but not in the orange and green dashed lines in Figure \ref{fig:m20_bt}, because cases with $B_{\rm i}=10^{12}$ G and $B_{\rm i}=10^{15}$ G end at phase a/b and d2, respectively, while the cases with $B_{\rm i}=10^{13}$ G and $B_{\rm i}=10^{14}$ G in short orbital period systems end at phase d2 and those in long orbital period systems end at phase a/b.
The second green circle ends at phase b while others end at phase a, which leads to a small bump in the green dashed line\footnote{This may be a numerical mistake because it jumps from phase a to phase b in the last step. Considering there is no significant influence on the results, we don't discuss it in the following.}.

We compare our results with two subclasses of HMXBs detected $P_{\rm s}$ and $P_{\rm orb}$, which are the supergiant X-ray binaries (SgXBs) and the supergiant fast X-ray transients (SFXTs).
The triangles and squares mark the SgXBs and the SFXTs respectively \footnote{data from \cite{Martinez-Nunez2017} and \cite{Tauris2017}}.
The SgXBs and SFXTs are two subclasses of HMXBs.
SgXBs are persistent systems in X-rays ($L_{\rm X} = 10^{36}-10^{38}$ erg s$^{-1}$), one of which consists of a compact accretor and a supergiant mass donor of spectral type O8-B1 I/II with strong stellar wind \citep{Martinez-Nunez2017,Tauris2017}.
While SFXTs are not persistent, which stay in quiescence at $L_{\rm X} = 10^{32}-10^{34}$ erg s$^{-1}$ for most of the time \citep{Romano2015} and exhibit short outbursts lasting a few hours reaching $10^{37}-10^{38}$ erg s$^{-1}$ \citep{Rampy2009,Bozzo2011}.
Figure \ref{fig:m20_bt} shows that most of the observed HMXBs, which have $P_{\rm s}<1000$ s and $P_{\rm orb}<30$ day, distribute in the blue region, meaning that they are probably normal neutron stars. 
Other five sources with $P_{\rm s}>1000$ s or $P_{\rm orb}>30$ day are in the red or orange regions, indicating that they may have strong magnetic fields.
One of them is in the orange region, which is probably a magnetar.
The red dashed line could be seen as the separatrix between magnetars and other NSs.

\begin{figure}
	\includegraphics[width=0.5\textwidth]{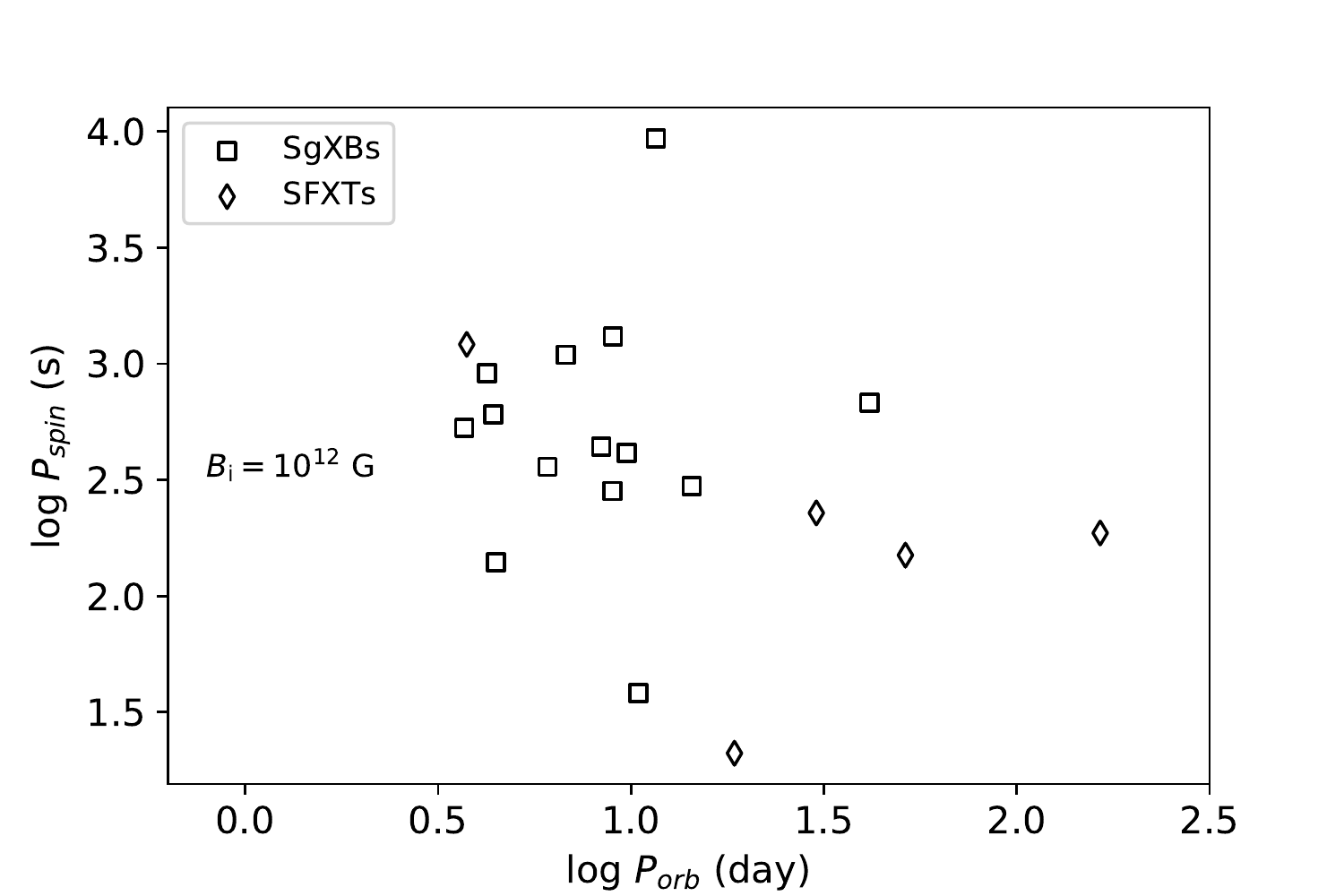}
	\includegraphics[width=0.5\textwidth]{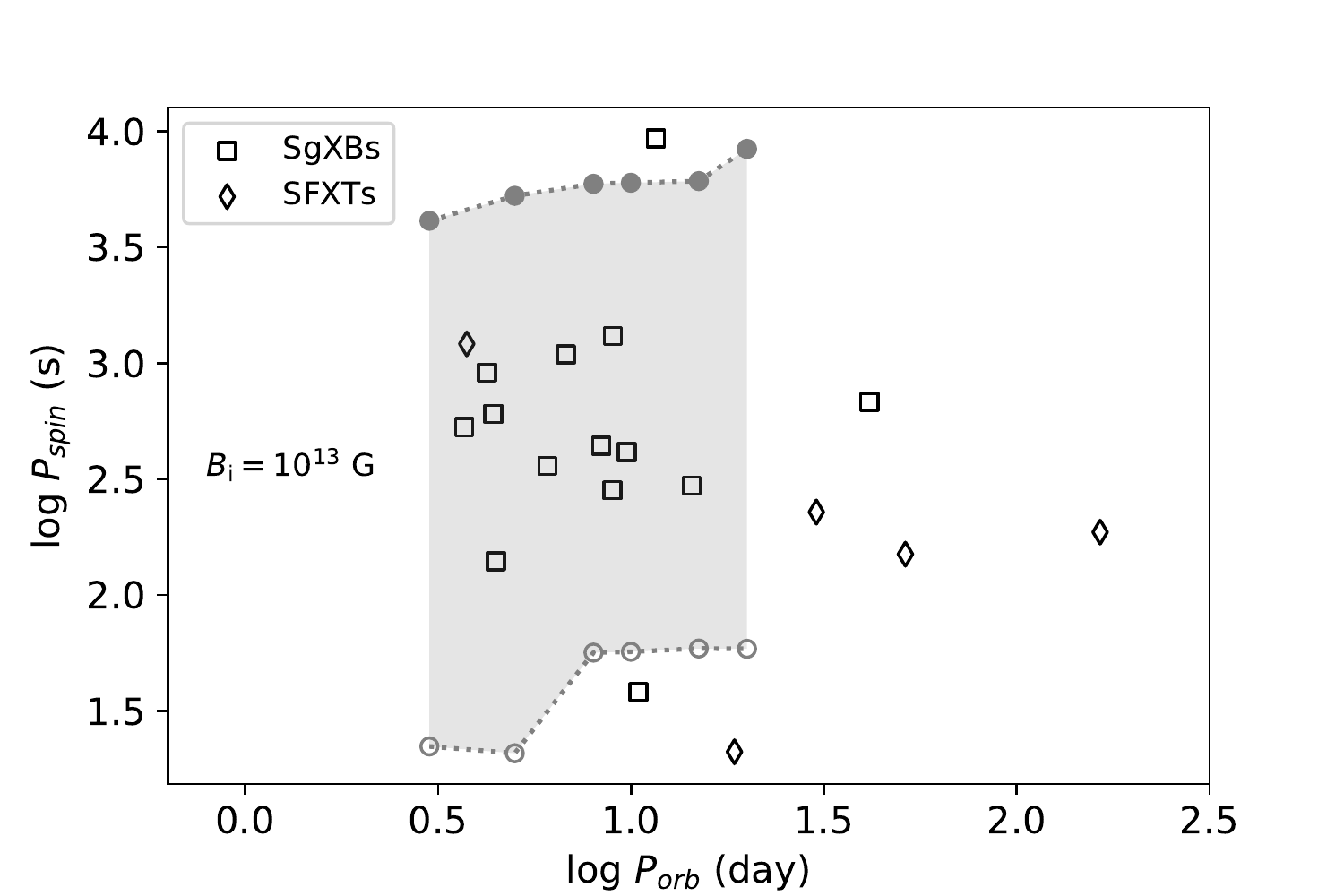}
	\includegraphics[width=0.5\textwidth]{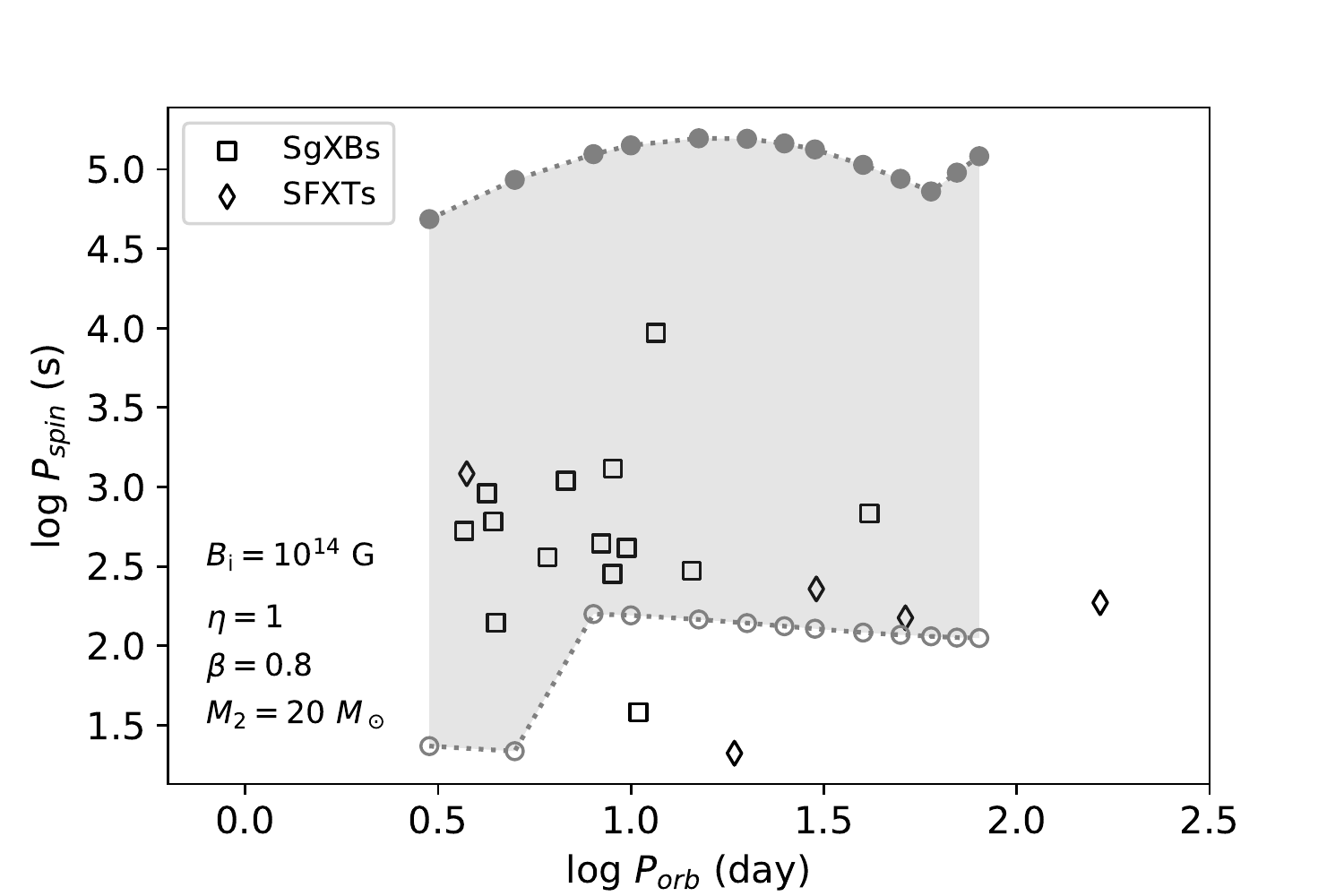}
	\includegraphics[width=0.5\textwidth]{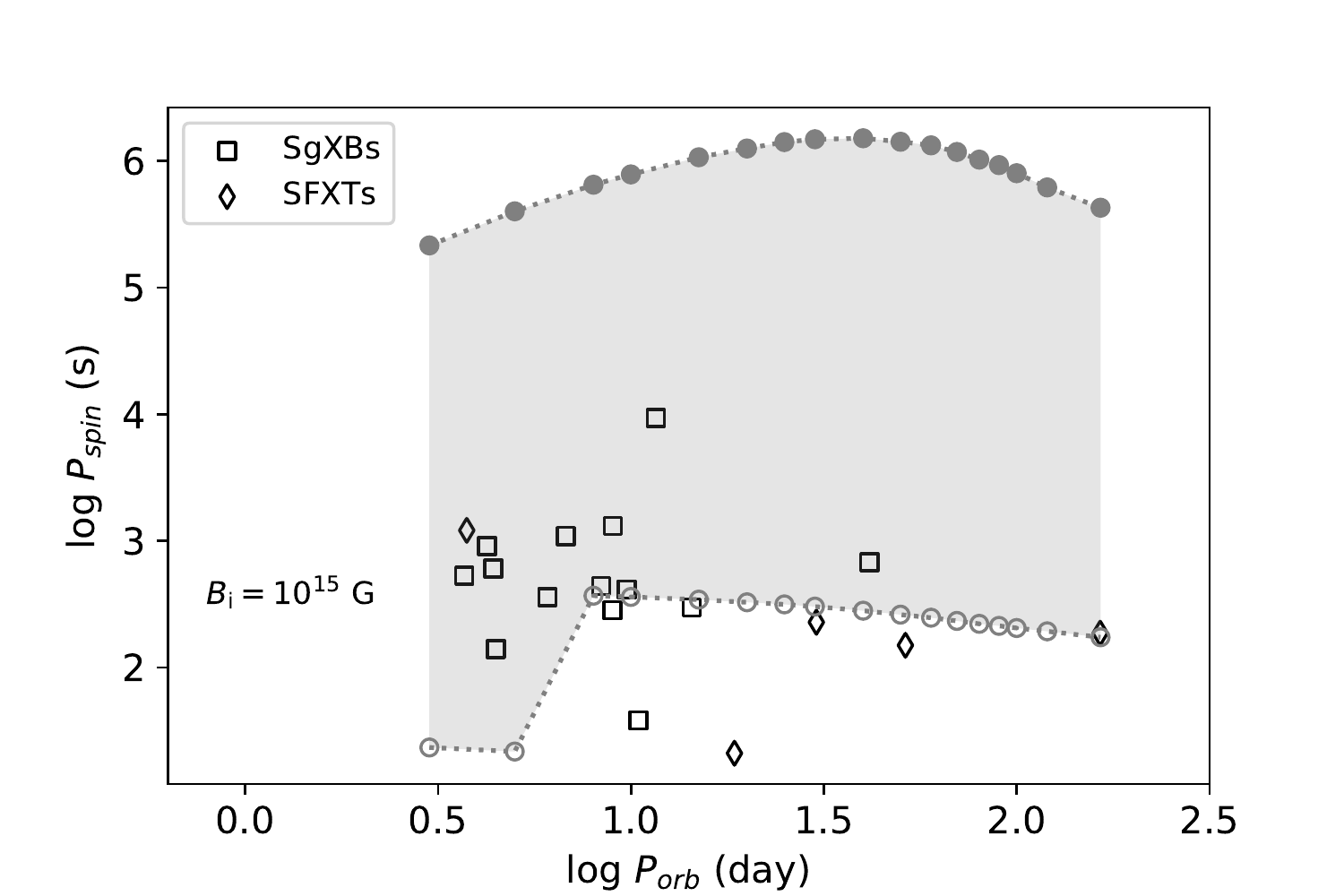}
	\caption{Results of the reference model with NSs' luminosity $L_{\rm X}>10^{32} {\rm ~erg ~s}^{-1}$ during the accretion phases. The gray filled and unfilled circles indicate the predicted maximum and minimum $P_{\rm s}$ of the NSs, respectively.
	And the gray regions give the predicted range of $P_{\rm s}$ of the NSs in each panel. From top-left to bottom-right panels, the magnetic fields are $B_{\rm i}=10^{12}$, $B_{\rm i}=10^{13}$, $B_{\rm i}=10^{14}$ and $B_{\rm i}=10^{15}$ G, respectively.
	There is no NS enters accretion phases in the cases of $B_{\rm i}=10^{12}$ G. }
	\label{fig:m20_bt_lx}
\end{figure}

Figure \ref{fig:m20_bt_lx} shows the $P_{\rm s}$ regions of NSs' luminosity $L_{\rm X}>10^{32} {\rm ~erg ~s}^{-1}$  during the accretion phases.
The gray filled and unfilled circles indicate the predicted maximum and minimum $P_{\rm s}$ of the NSs, respectively.
And the gray regions give the predicted range of $P_{\rm s}$ of the NSs in each panel. From top-left to bottom-right panels, the magnetic fields are $B_{\rm i}=10^{12}$, $B_{\rm i}=10^{13}$, $B_{\rm i}=10^{14}$ and $B_{\rm i}=10^{15}$ G, respectively.
In the cases of $B_{\rm i}=10^{12}$ G, no NS enters accretion phases.

We plot the spin evolution process of the reference model with $P_{\rm orb} = 10$ day in Figure \ref{fig:m20_po10}. 
From lower to upper, the dotted, dashed, dot-dashed and solid lines represent the evolution of NSs with magnetic field $B_{\rm i}=10^{12}$, $B_{\rm i}=10^{13}$, $B_{\rm i}=10^{14}$ and $B_{\rm i}=10^{15}$ G, respectively. 
In each line, the black, orange, green, red and blue parts indicate that the NSs are in phase a, b, c, d1 and d2.
The right panel is the magnified part of the left panel with $5<{\rm log} t ~ ({\rm yr})<7$.

\subsection{Parameter Study}

The main configurable parameters in our model are the companion mass $M_2$, the wind parameters $\eta$ and $\beta$.
In this subsection, we vary these parameters based on the reference model to see how they can influence the results.

\subsubsection{The Companion Mass $M_2$}

\begin{figure}[ht!]
	\centering
	\includegraphics[width=0.5\textwidth]{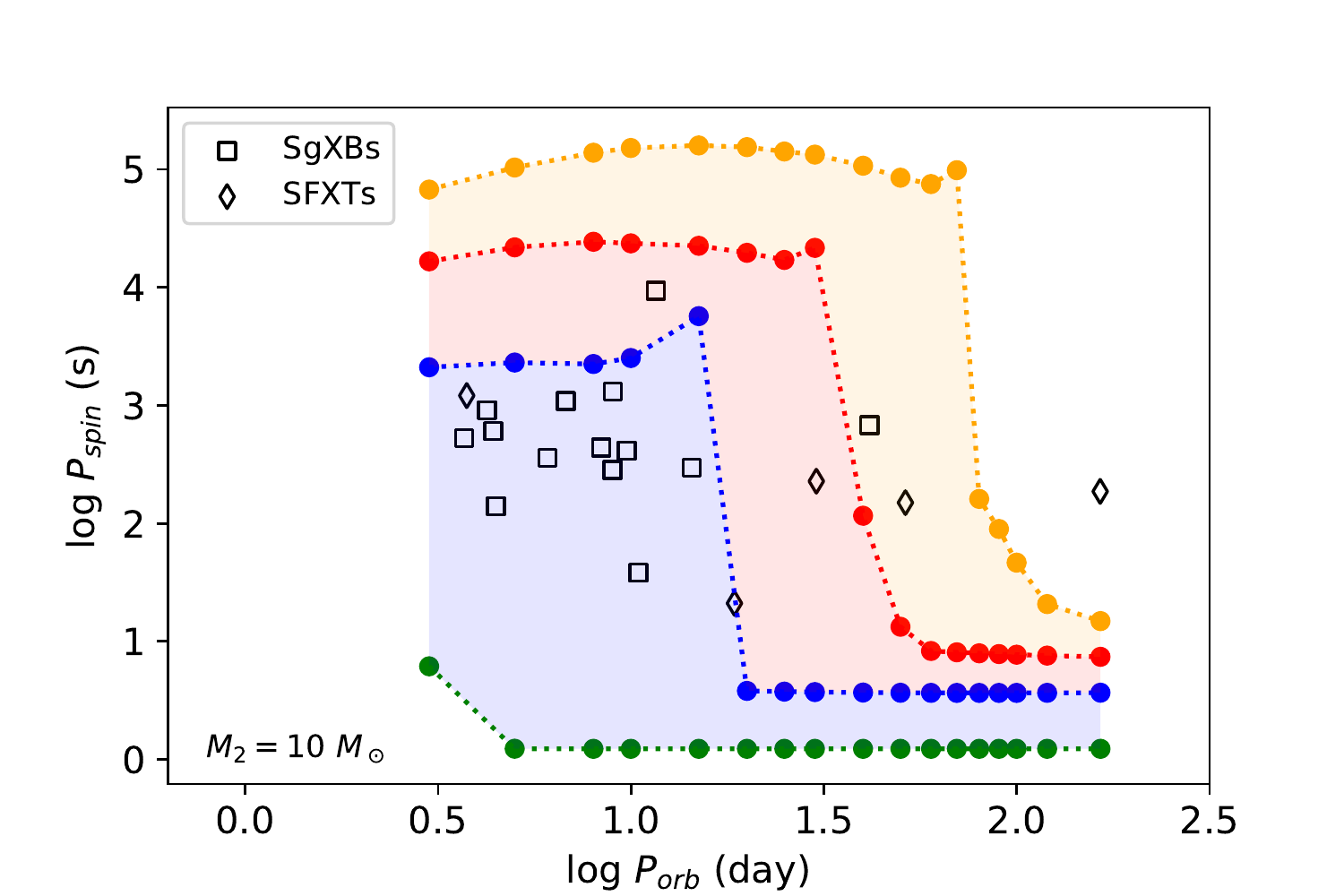}
	\includegraphics[width=0.5\textwidth]{pp_m20.pdf}
	\includegraphics[width=0.5\textwidth]{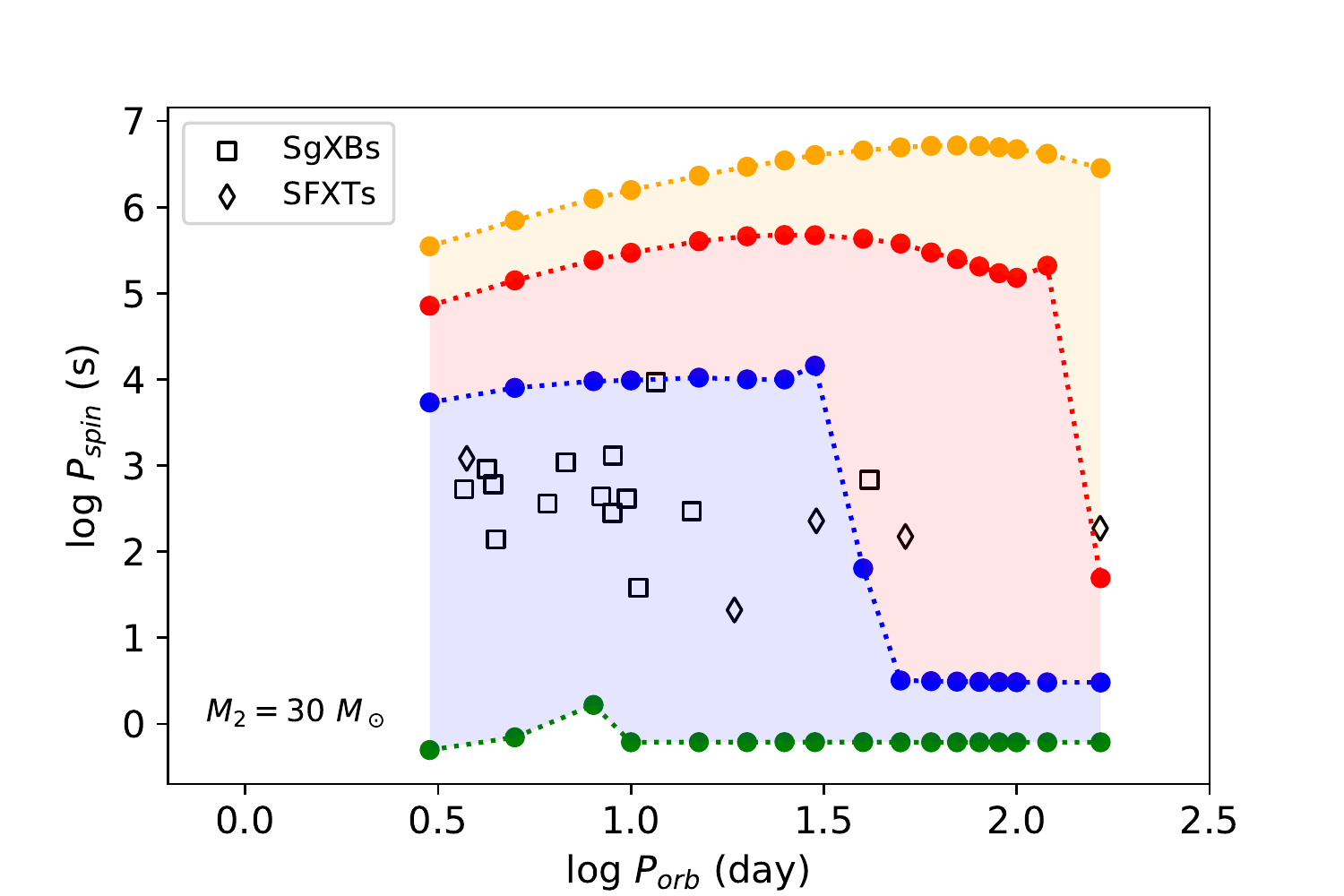}
	\caption{Same as Figure \ref{fig:m20_bt} but with the companion mass $M_2=$ 10, 20 (the reference model) and 30, from top to bottom. }
	\label{fig:m_bt}
\end{figure}

We first change the mass of the companion $M_2$.
The results of maximum spin period with different companion mass are exhibited in Figure \ref{fig:m_bt}. From top to bottom, $M_2$ is taken to be $10 ~ M_\odot$, $20 ~ M_\odot$ (the reference model) and $30 ~ M_\odot$.
It shows that more sources fall in the red and orange regions in the upper panel while less sources in these regions in the lower panel compared with the reference model.
This may imply that systems with less massive companions are more likely to dedicate HMXBs with magnetars.

\subsubsection{The Wind Coefficient $\eta$}

\begin{figure}
	\includegraphics[width=0.5\textwidth]{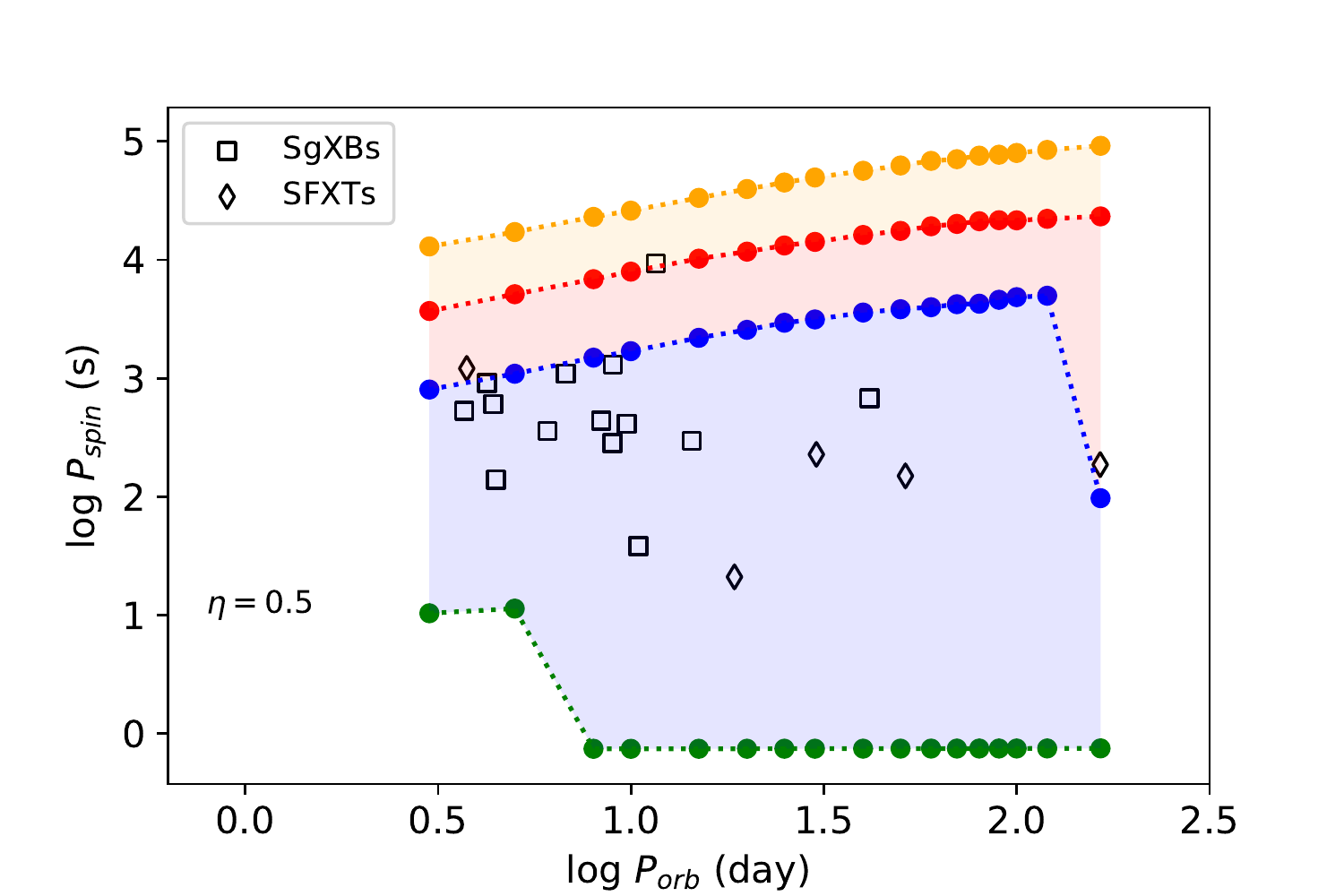}
	\includegraphics[width=0.5\textwidth]{pp_m20.pdf}
	\includegraphics[width=0.5\textwidth]{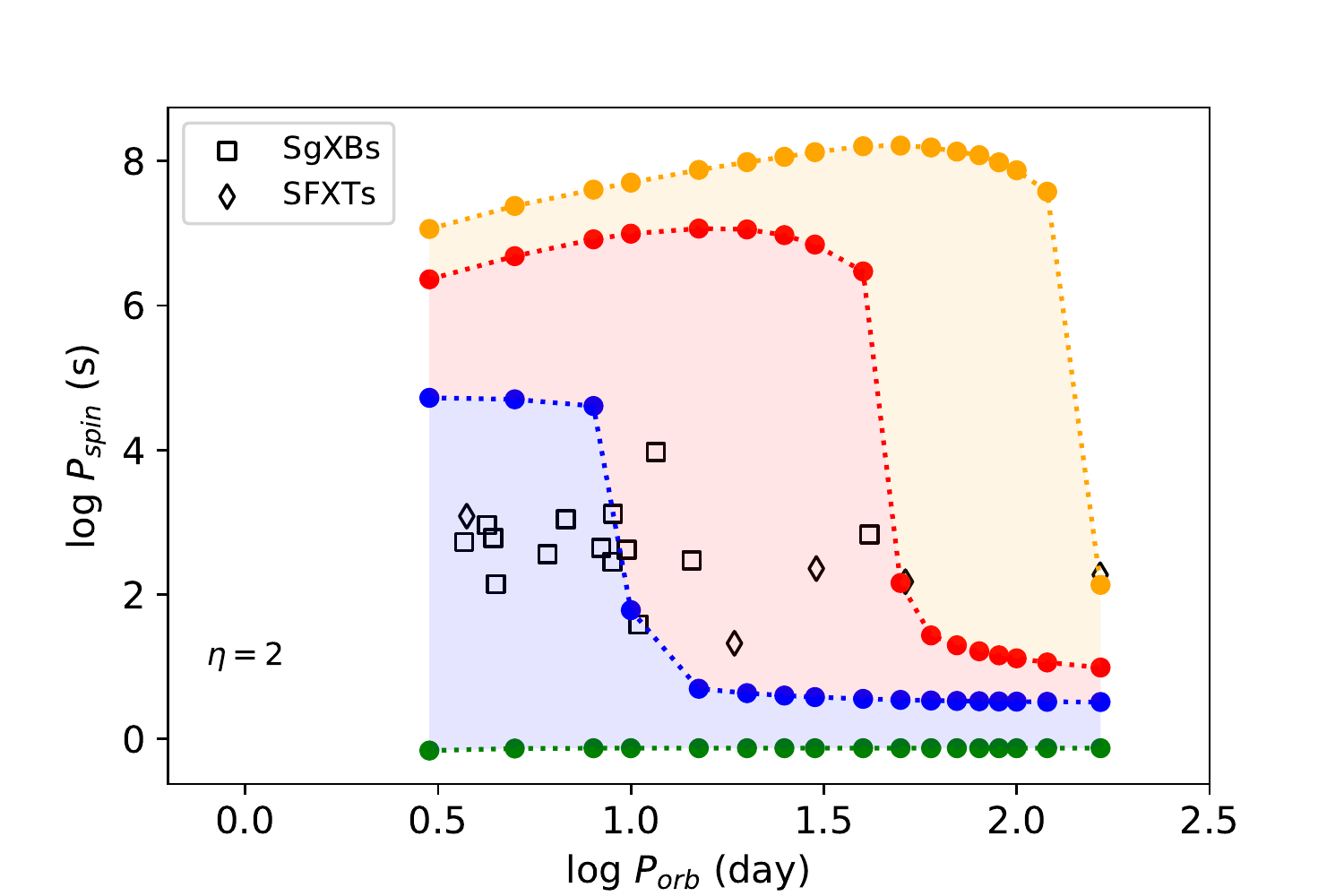}
	\includegraphics[width=0.5\textwidth]{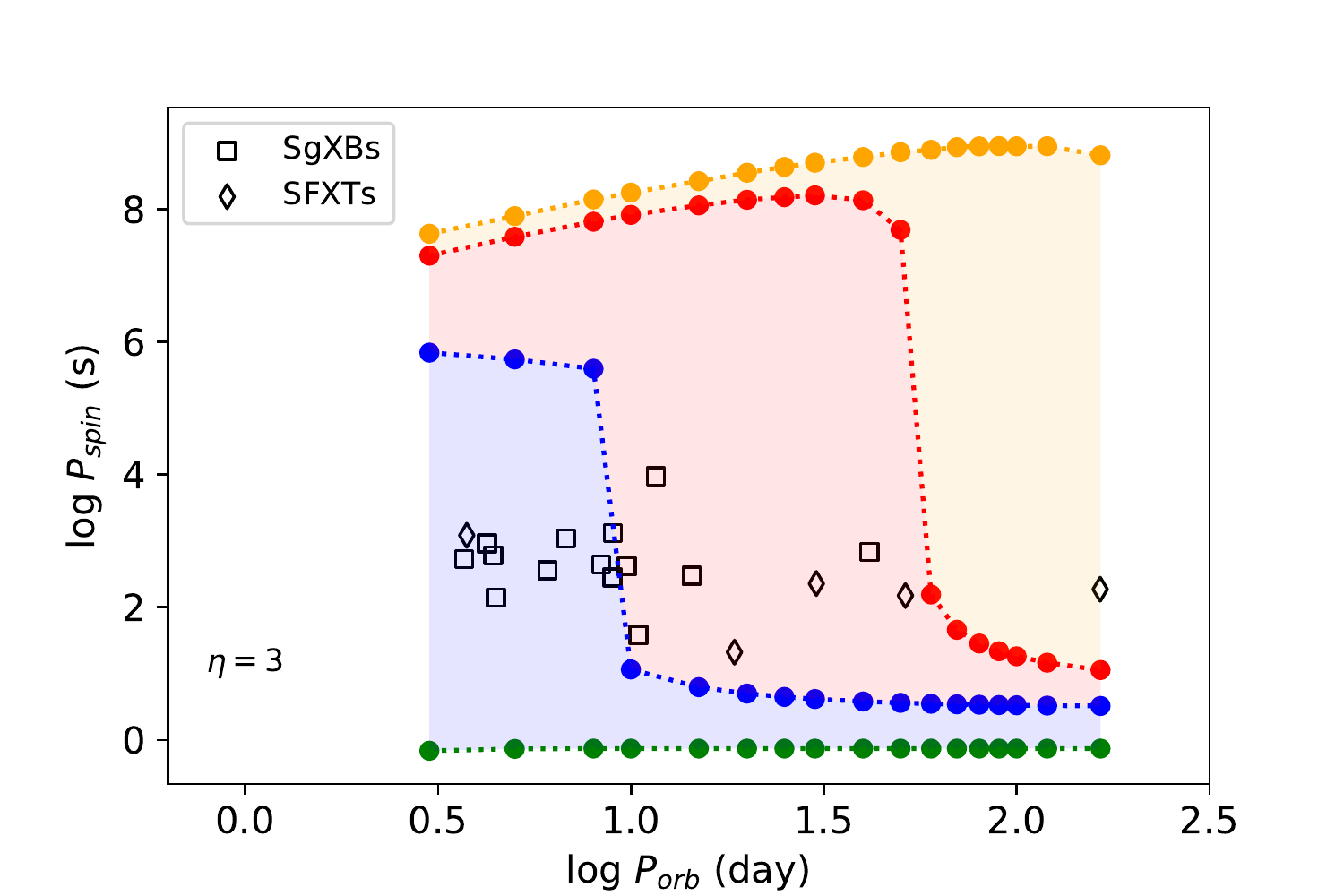}
	\caption{Same as Figure \ref{fig:m20_bt} but with the wind coefficient $\eta=$ 0.5, 1 (the reference model), 2 and 3 from top-left to bottom-right. Comparing with the reference model, there are less sources in the red and orange regions when $\eta = 0.5$, while more sources in these regions when $\eta  = 2$ or 3. This may imply that the larger the wind coeﬀicient, the more magnetars are in HMXBs. }
	\label{fig:eta_bt}
\end{figure}

Figure \ref{fig:eta_bt} shows the dependence of $P_{\rm max}$ on the wind coefficient $\eta$, where $\eta=0.5$, 1 (the reference model), 2 and 3 from top-left to bottom-right.
Comparing with the reference model, we find that less sources in the red and orange regions when $\eta=0.5$, while more sources in these regions when $\eta=$ 2 or 3. 
This may imply that the larger the wind coefficient, the more magnetars are in HMXBs.

\subsubsection{The Wind Power Law Index $\beta$}

\begin{figure}
	\includegraphics[width=0.5\textwidth]{pp_m20.pdf}
	\includegraphics[width=0.5\textwidth]{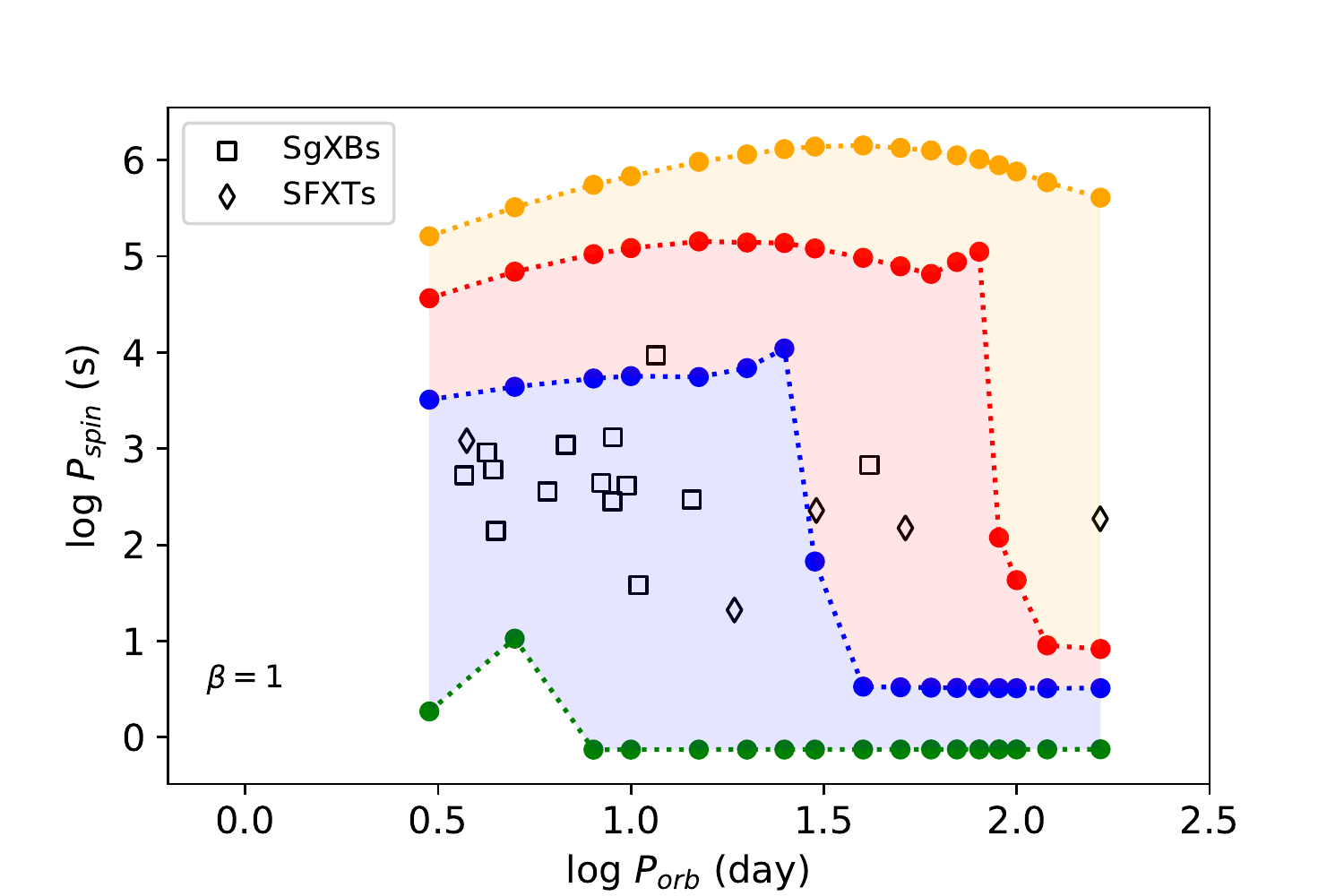}
	\includegraphics[width=0.5\textwidth]{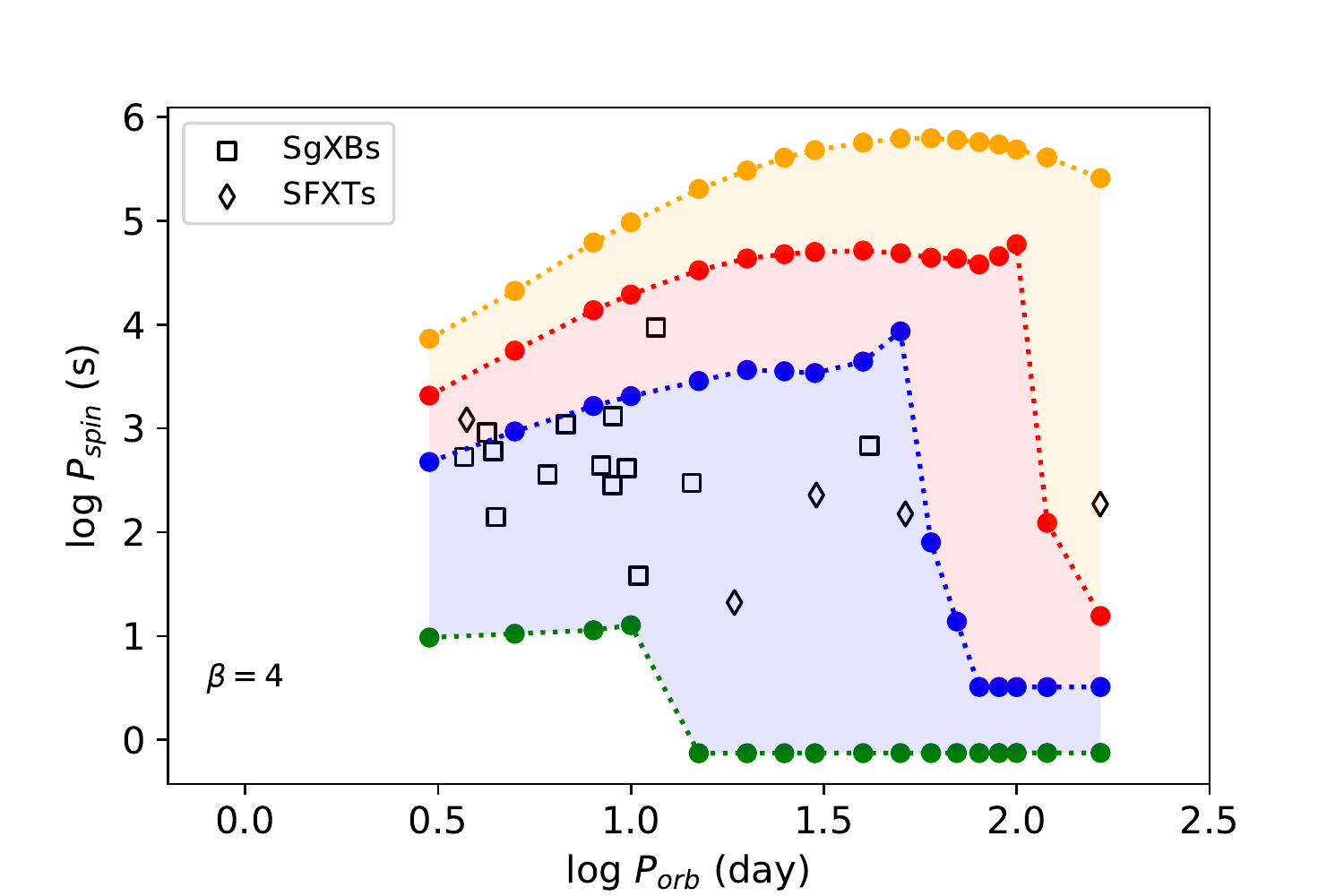}
	\includegraphics[width=0.5\textwidth]{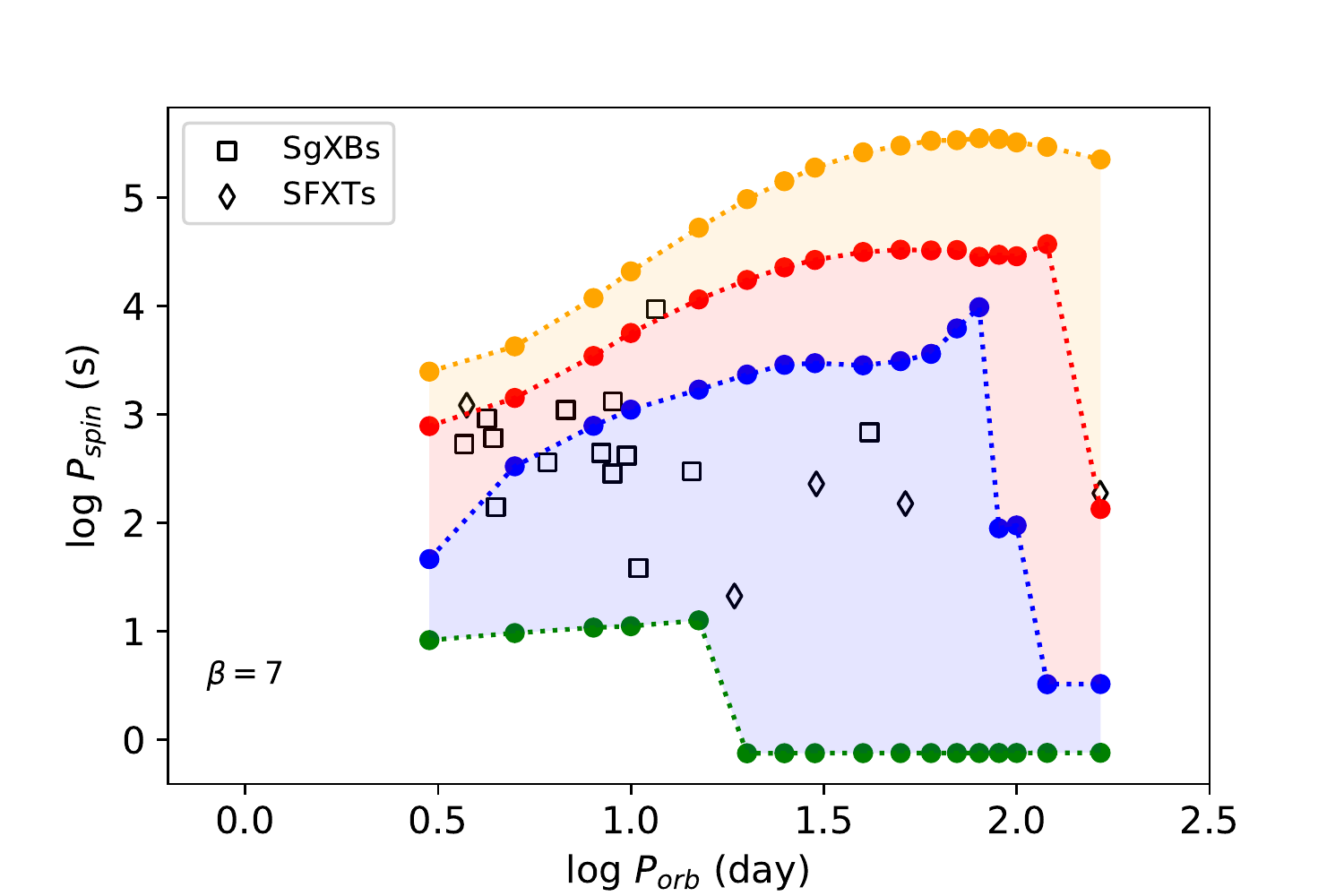}
	\caption{Same as Figure \ref{fig:m20_bt} but with the wind power index $\beta=$ 0.8 (the reference model), 1, 4 and 7 from top-left to bottom-right. It shows that when $\beta$ is much larger, some sources with short orbital periods need stronger magnetic fields to explain their long spin periods.}
	\label{fig:beta_bt}
\end{figure}

\cite{Karino2020} discussed the influence of parameter $\beta$ within larger scale, where $\beta=1$ and $7$ indicate the fast and slow wind cases, respectively. They found that the NSs spin down rapidly in the slow wind cases due to the propeller effect and settling accretion shell, while magnetic inhibition causes spin-down in the fast cases.
Figure \ref{fig:beta_bt} presents the $P_{\rm max}$ results with different values of wind power law index $\beta$. From top-left to bottom-right, we take $\beta= 0.8$ (the reference model), 1, 4 and 7, respectively. It shows that when $\beta$ is much larger, some sources with short orbital periods need strong magnetic fields to explain their long spin periods. But we are not sure whether  $\beta$ can be as large as 7.
This may indicate that in the slow wind cases \citep{Karino2020}, {an NS needs} a strong magnetic field to get a large $R_{\rm lc}$ to start the interaction with the wind material.

\section{Discussion}

\subsection{Special Sources}

\subsubsection{J11215$-$5952}

From Figures \ref{fig:m20_bt}, \ref{fig:eta_bt} and \ref{fig:beta_bt}, we know that only one source is always in the red or orange regions, which is J11215$-$5952 (hereafter J11215) with orbital period $P_{\rm orb} = 164.6$ day and spin period $P_{\rm s} = 186.78$ s \citep{Sidoli2006,Sidoli2007,Sidoli2020,Masetti2006}.
That means J11215 needs an initial magnetic field $B_{\rm i}>10^{13}$ G to explain its orbital and spin period, which implies that it's probably a magnetar.
In a wide-separation binary, the NS may need a large effective radius to let the wind be captured.
So J11215 needs a strong magnetic field to get a long spin period as well as a large $R_{\rm lc}$ in phase a. 

\subsubsection{OAO 1657$-$415 and J18483$-$0311}
\begin{figure}
	\includegraphics[width=0.5\textwidth]{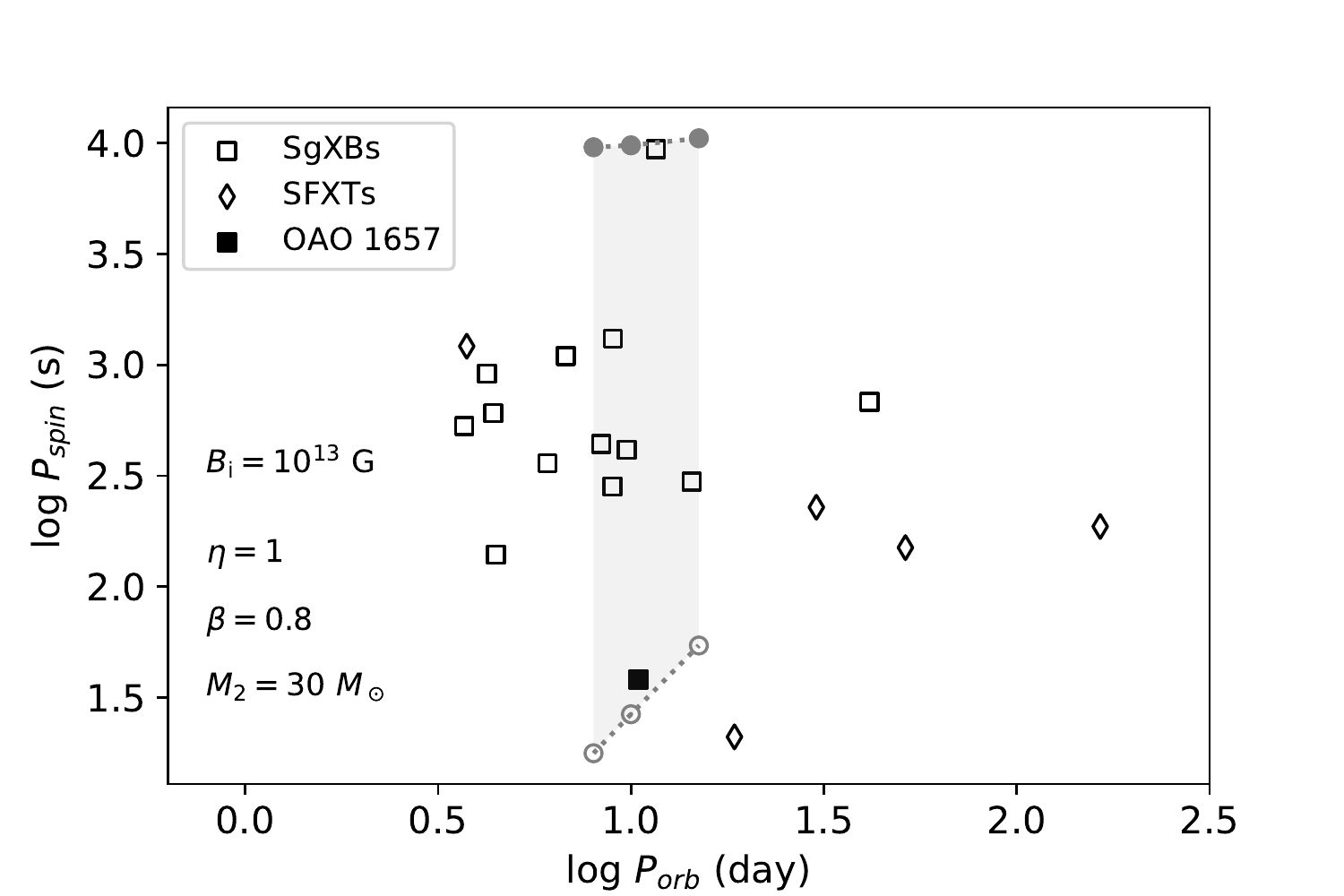}
	\includegraphics[width=0.5\textwidth]{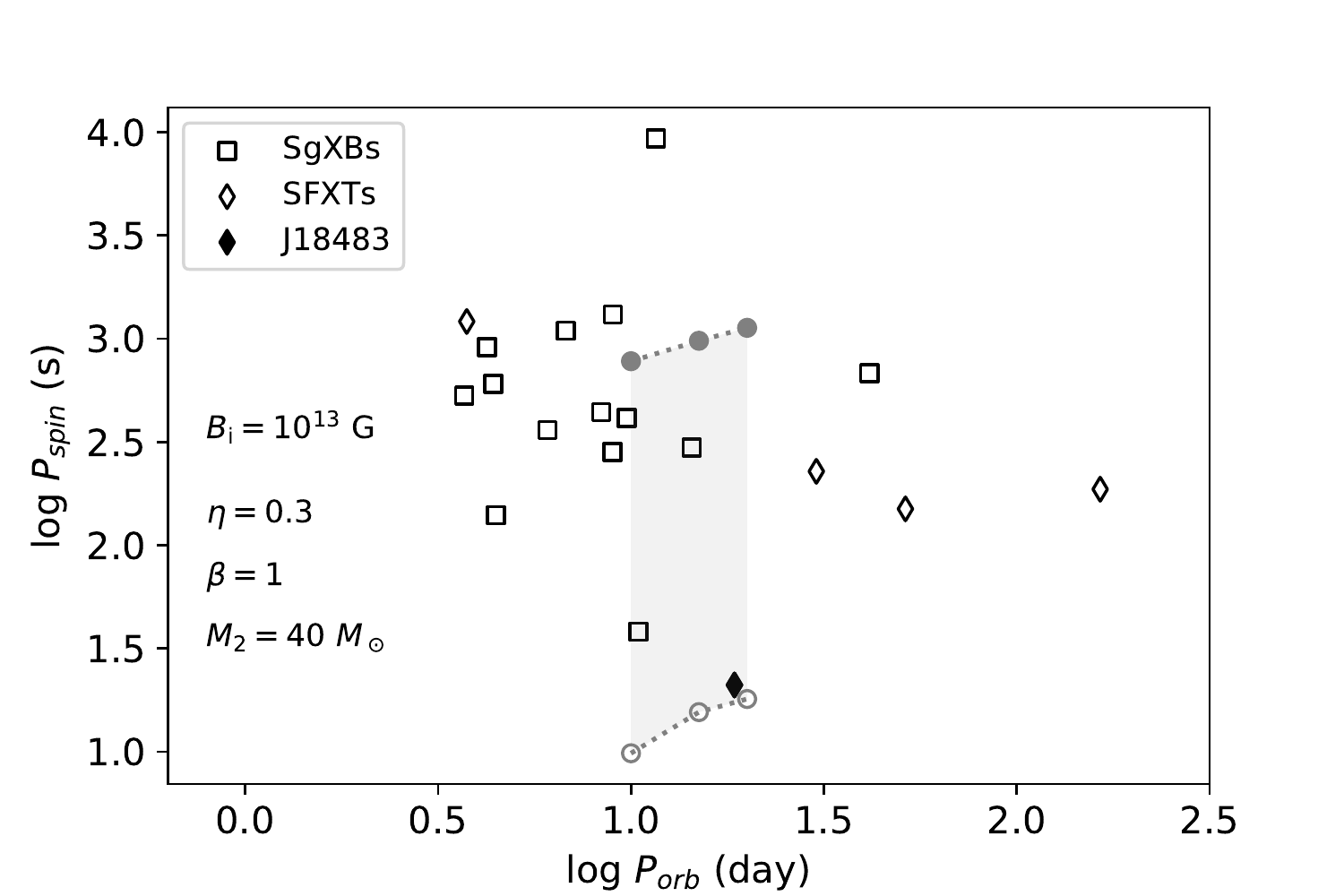}
	\caption{Parameters that can cover OAO 1657 (the left panel) and J18483 (the right panel).}
	\label{fig:OAO_1657}
\end{figure}

There are two sources not covered in Figure \ref{fig:m20_bt_lx} because their spin periods are too small,
which are OAO 1657$-$415 (hereafter OAO 1657) with $P_{\rm orb} = 10.448$ day and $P_{\rm s} = 38.2$ s and J18483$-$0311 (hereafter J18483) with $P_{\rm orb} = 18.55$ day and $P_{\rm s} = 21$ s.
We think these sources are or have been in directly accretion phase (phase d1) so that they can spin up to a very fast rotation.
We show two groups of parameters that can cover OAO 1657 and J18483 in Figure \ref{fig:OAO_1657}, in which we can explain OAO 1657 within the parameters space in our model while J18483 needs a small $\eta$ and massive companion.

\subsubsection{4U 1954$+$31}

\begin{figure}
	\centering
	\includegraphics[width=0.5\textwidth]{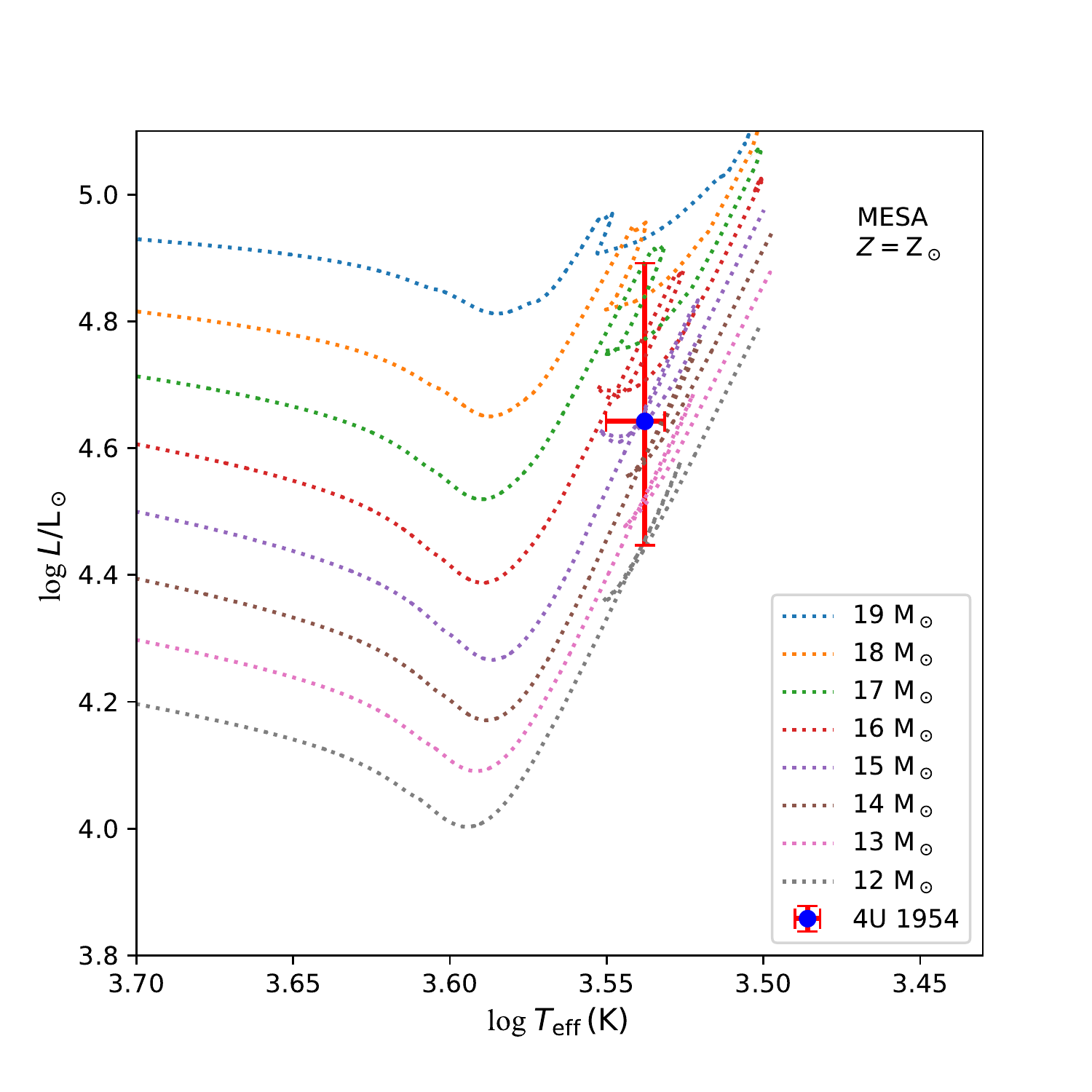}
	\includegraphics[width=0.5\textwidth]{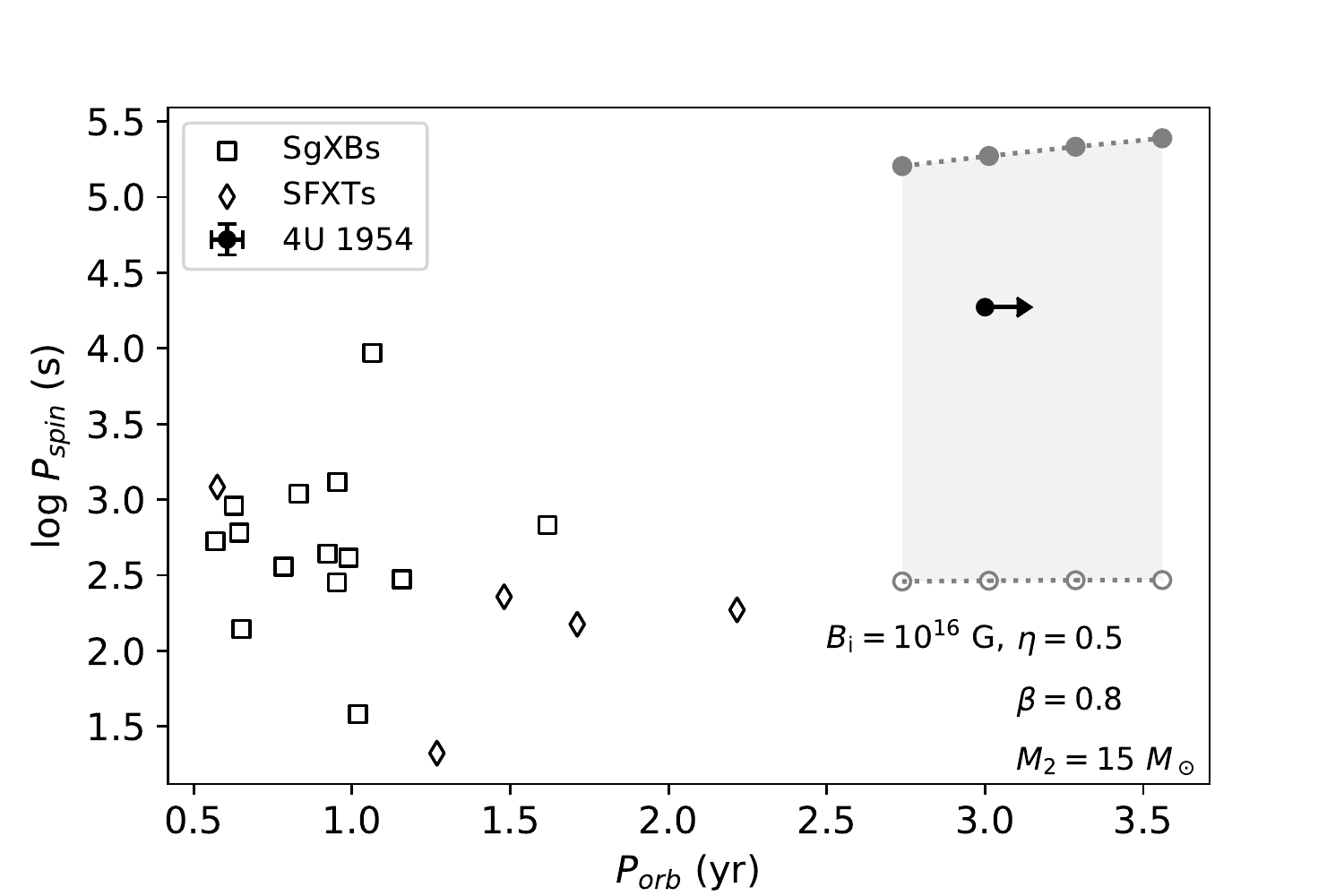}
	\caption{Upper: Location of the companion of 4U 1954 on the HRD.
			Lower: A group of parameters that can cover 4U 1954, in which the initial magnetic field is $B_{\rm i}=10^{16}$ G.}
	\label{fig:4u1954}
\end{figure}

Recently, 4U 1954$+$31 has been re-classified as a HMXB containing a late-type supergiant with mass of $9^{+6}_{-2} ~M_\odot$ \citep{Hinkle2020}.
While the companion mass arrived according to its position in the Hertzsprung–Russell Diagram (HRD) is model depended, as there are still many uncertainties plague the evolutionary tracks of stars, for instance the stellar wind and mixing. As a comparison, we also evolve some models using the stellar evolutionary code Modules for Experiments in Stellar Astrophysics (MESA) \citep{Paxton2011} and compare their evolutionary tracks with 4U1954+31 in HRD (the upper panel of Figure \ref{fig:4u1954}). In our models, the convection is calculated following the standard mixing-length theory \citep{Bohm1958} with the scale parameter $\alpha_{\rm MLT}=1.5$, and the convective regions are determined by the Ledoux criterion. The semi-convection parameter are assumed to be $\alpha_{\rm SC}=1.0$. Besides, the mass loss due to stellar wind are calculated according to their effective temperature and surface Hydrogen (H) mass fraction. That is, the prescriptions of \cite{Vink2001} and \cite{Nugis2000} are used for hot H-rich ( $T_{\rm eff}>10^4\rm K$ and $X_{\rm s} \geq 0.4$) and H-poor ($T_{\rm eff}>10^4\rm K$ and $X_{\rm s} < 0.4$) stars, respectively. While for $T_{\rm eff}<10^4\rm K$ we change to that of \cite{deJager1988}. We simulate their evolution from zero age main sequence (ZAMS) to core-Carbon (C) ignition, with initial mass $M_{\rm i}$ ranging from 12 to 19 $M_{\odot}$. Our predicted mass of 4U 1954+31 is $15^{+3}_{-3}\rm M_{\odot}$, which is $\sim 6\rm M_{\odot}$ heavier than that of \cite{Hinkle2020}.
The spin period of the NS is $\sim 5$ hr \citep{Corbet2008,Marcu2011,Enoto2014} while the orbital period is not confirmed.
Only a lower limit of 3 yrs is given \citep{Hinkle2020}, which is much longer than the orbital periods of SgXBs and SFXTs in Figure \ref{fig:m20_bt}.
And we can't explain it within the parameter space used above, but the results in the lower panel of Figure \ref{fig:4u1954} can cover it with the following parameters: $\eta=0.5$, $\beta=0.8$, $B_{\rm i}=10^{16}$ G and $M_2=15 ~M_\odot$.
Since the separation of this system is very wide, another possible model to explain its long spin period is that the NS is firstly spun down by a fallback disk and then interacts wind material.

\subsection{The distribution of NSs in HMXBs}

\begin{figure}
	\centering
	\includegraphics[width=0.5\textwidth]{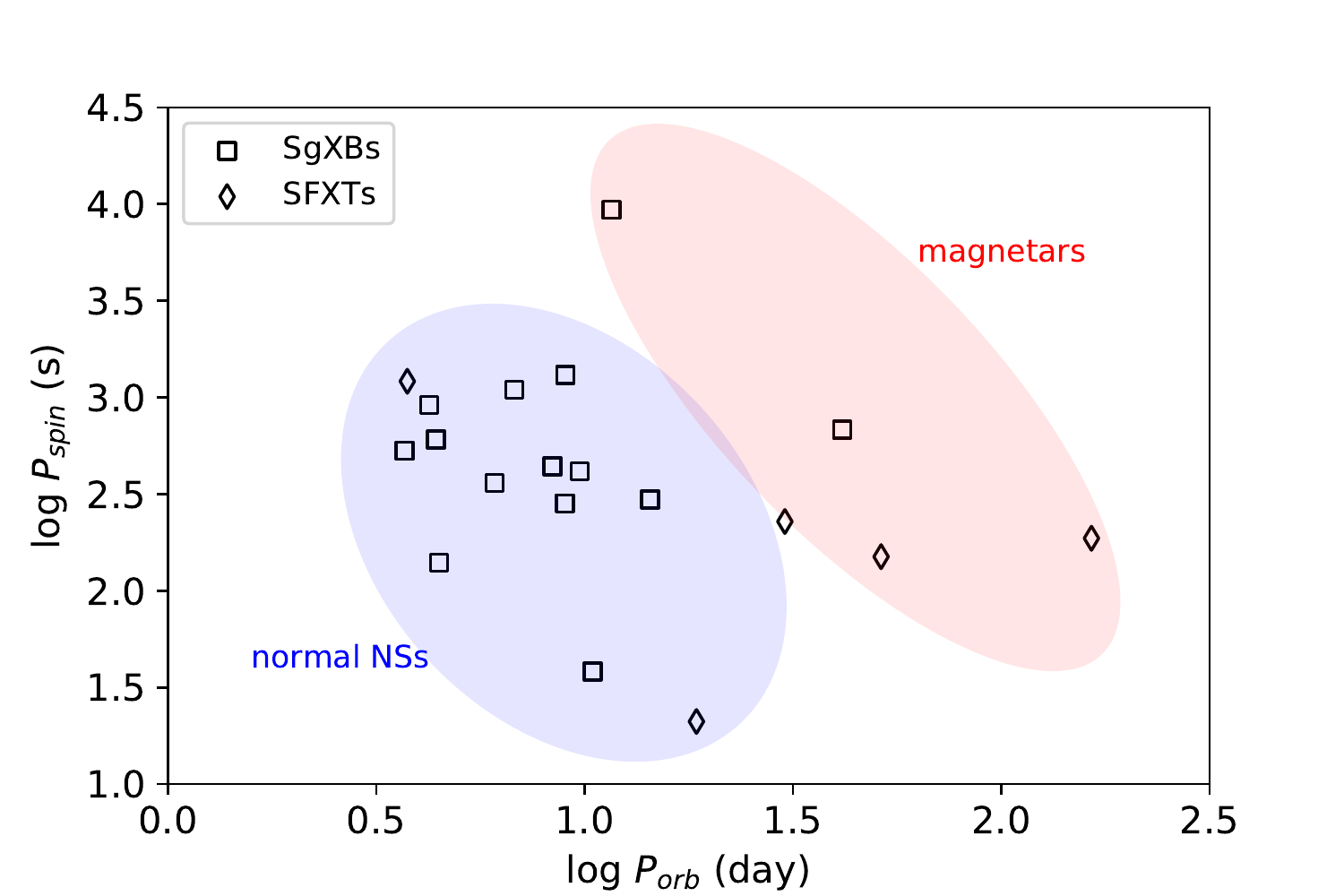}
	\caption{The distribution of magnetars and normal NSs in HMXBs predicted by our model.}
	\label{fig:magnetar_predict}
\end{figure}

Figure \ref{fig:magnetar_predict} depicts the predicted distribution of magnetars and normal magnetic field NSs in HMXBs.
In the Corbet diagram, the normal NSs may center at the region where the spin periods and orbital periods are both short, while magnetars may locate at the region where either $P_{\rm s}$ or $P_{\rm orb}$ are long.
As proposed in \cite{Li1999}, the reason is that a strong magnetic field makes a NS achieving a slower spin period in phase a by magnetic dipole radiation, as well as a larger effective radius, so that the NS is easy to capture the wind stellar material and enters the interaction phases (phase b, c ,d1 and d2) earlier, then the interaction between the wind material and the magnetosphere could have much time to spin the NS down.

\section{Summary}

In this work, we explore the possible parameter space for magnetars in high-mass X-ray binaries by performing spin evolution of neutron stars. 
We arrange and simplify the wind accretion regimes taking the direct and the subsonic settling accretion into account.
Compared with previous studies, we use grid methods combining the individual and population synthesis method and simulate a more complete evolution process, including the evolution of winds from the companions and the decay of the magnetic fields of the NSs.
Our results show that some NSs in the right and upper region in the Corbet diagram, which are in wide-separation systems or have long spin periods, need magnetic fields in magnetar magnitude to reproduce their relatively long spin periods. This implies that some NSs may born as magnetars in HMXBs, then their magnetic fields decay to the order of normal NSs through a few Myrs, but the spin periods record their evolution informations and can provide some traces about their initial magnetic fields.

In the calculations we employ the magnetic field evolution model modified by \cite{Fu2012}, where the parameters are taken as $\alpha=1.6$ and $\tau_{\rm d,i}=10^{3} {\rm ~yr}/(B_{\rm i}/10^{15} {\rm ~G})^{\alpha}$. There are some debates on these parameters \citep{DallOsso2012,Fu2012,Gullon2014} and we study them in the appendix.
The simple $\beta$-velocity law used in our work may not accurately describe the wind profile of the companion since some self-consistent calculations indicate that the wind velocity field is more complicated \citep{Grafener2005,Grafener2008,Sander2015}.
We model circular orbits only in the calculations while most of the HMXBs have eccentric orbits because the NSs receive a natal kick at the moment of their formation.
The separation becomes variable in eccentric orbits, which leads to parameters more complicated, including the wind velocity, the wind density, the accretion rate, the NS orbital velocity and so on. 
Because the orbital period is far shorter than the lifetime of the system, so for simplicity, we can use a circular orbit to replace the influence of an eccentric orbit and there is no significant impact on the results.

\normalem
\begin{acknowledgements}
	We thank the anonymous referee for helpful comments and suggestions. 
	This work was supported by the Natural Science Foundation of China under grant No. 11933004, 11988101, 12041301, 12063001, 11773015, Project U1838201 supported by NSFC and CAS, the program A for Outstanding PhD candidate of Nanjing University, the National Key Research and Development Program of China (2016YFA0400803) and the Fundamental Research Funds for the Central Universities.
\end{acknowledgements}

\appendix

\section{Parameter study of the evolution of the magnetic field}

\begin{figure}
	\includegraphics[width=0.5\textwidth]{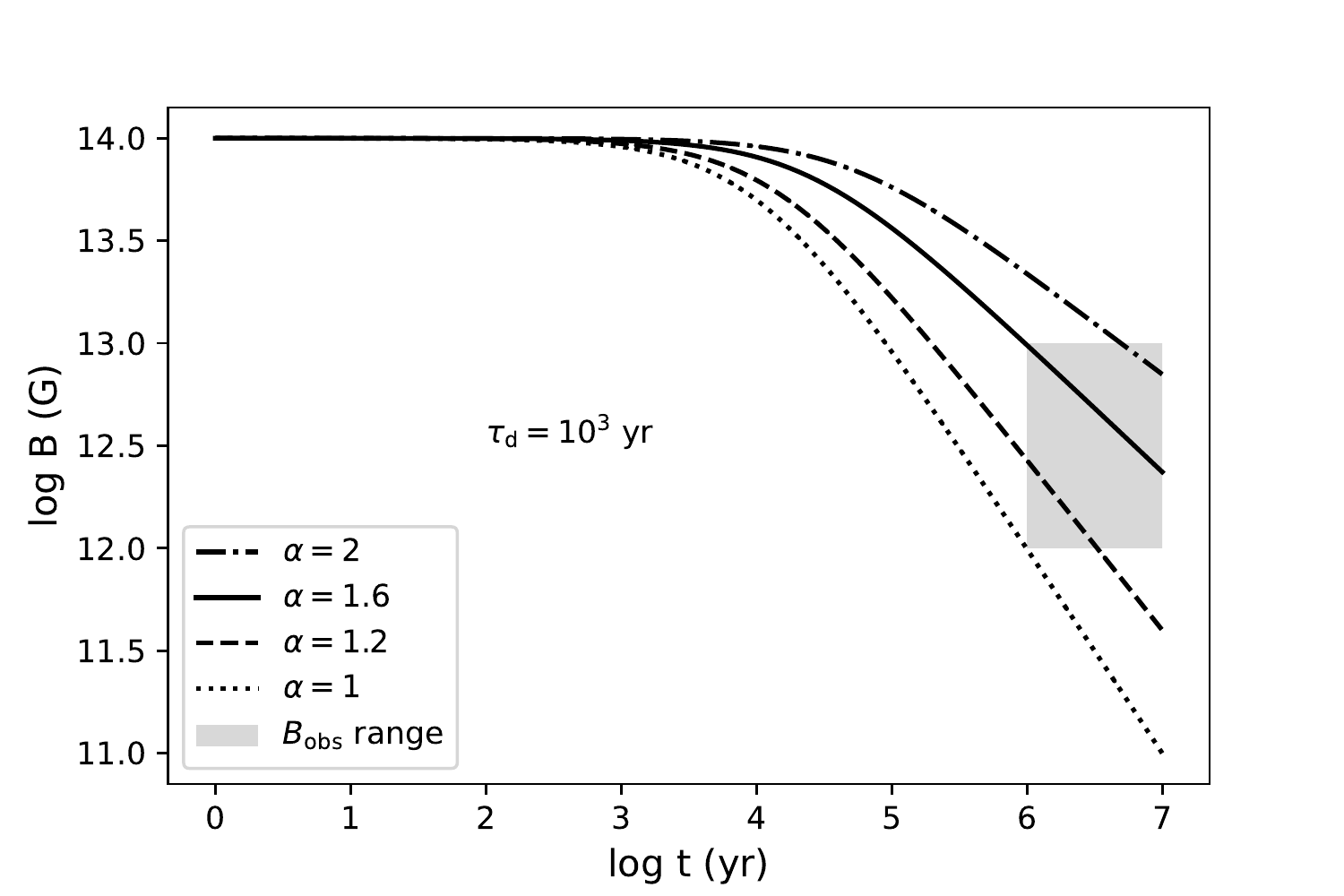}
	\includegraphics[width=0.5\textwidth]{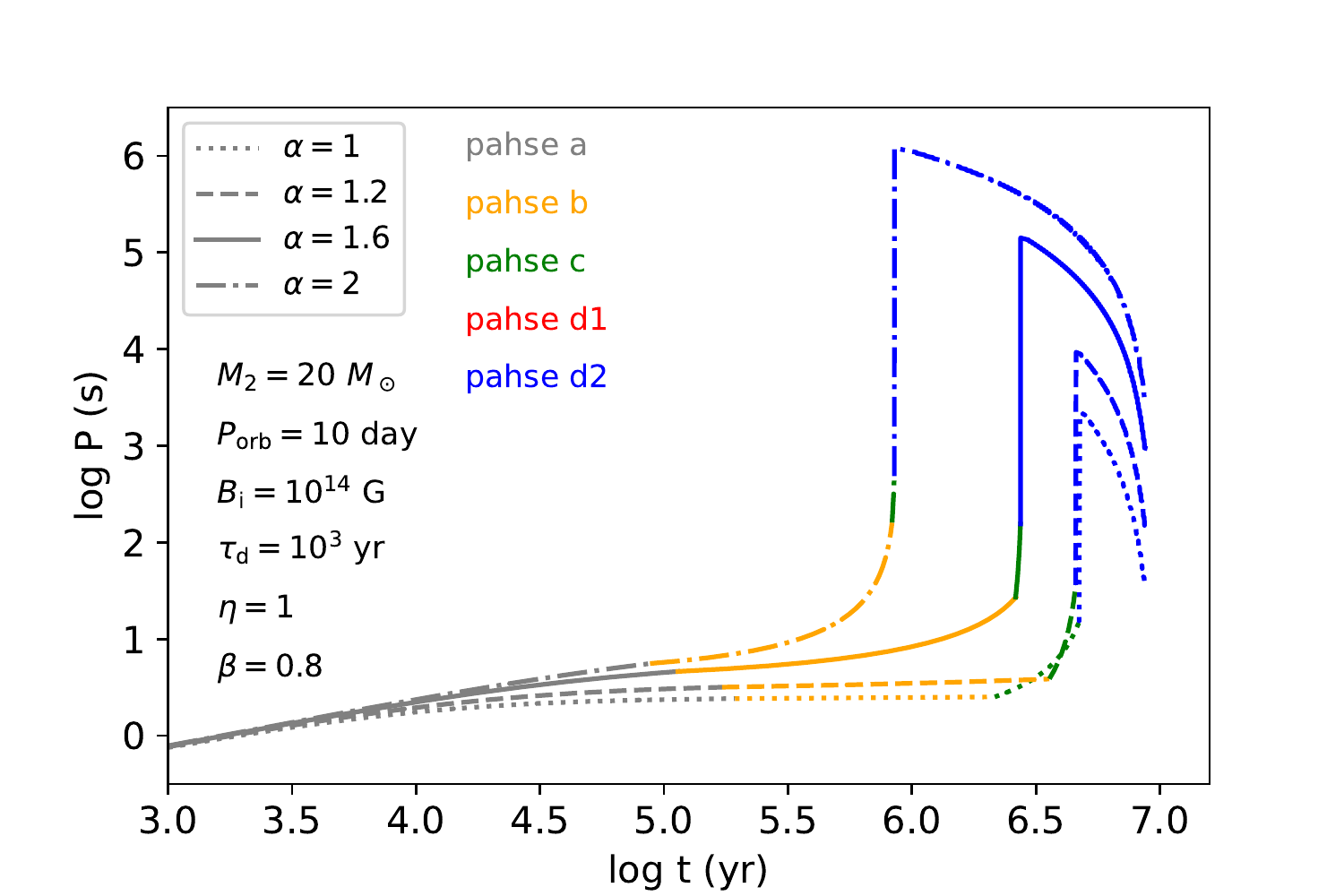}
	\caption{The magnetic field (left panel) and spin (right panel) evolution tracks with different $\alpha$. The gray region in the left panel covers the range of the estimated magnetic field and age of NSs in HMXBs from observetions.}
	\label{fig:BP_evo_alpha}
\end{figure}

\begin{figure}
	\includegraphics[width=0.5\textwidth]{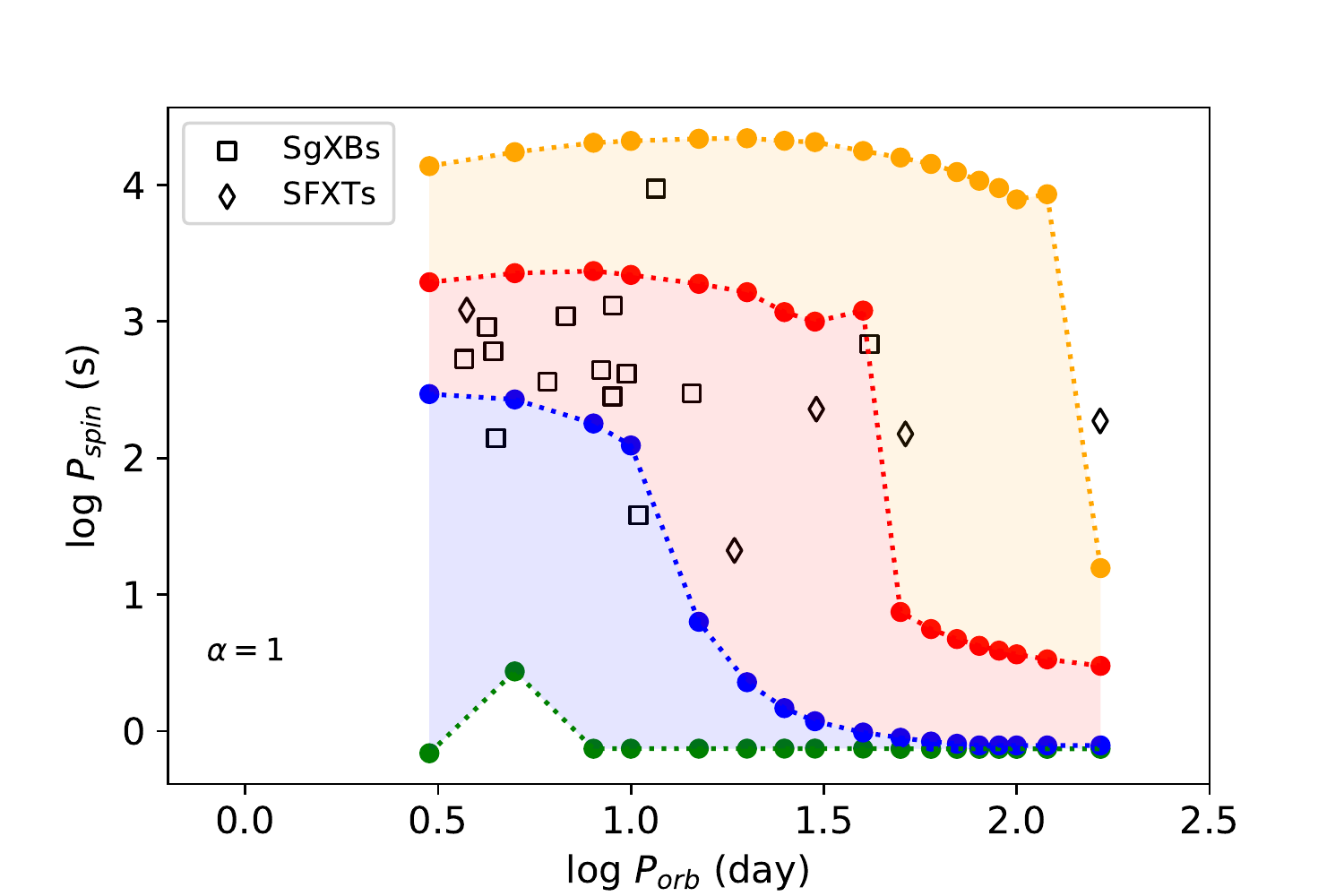}
	\includegraphics[width=0.5\textwidth]{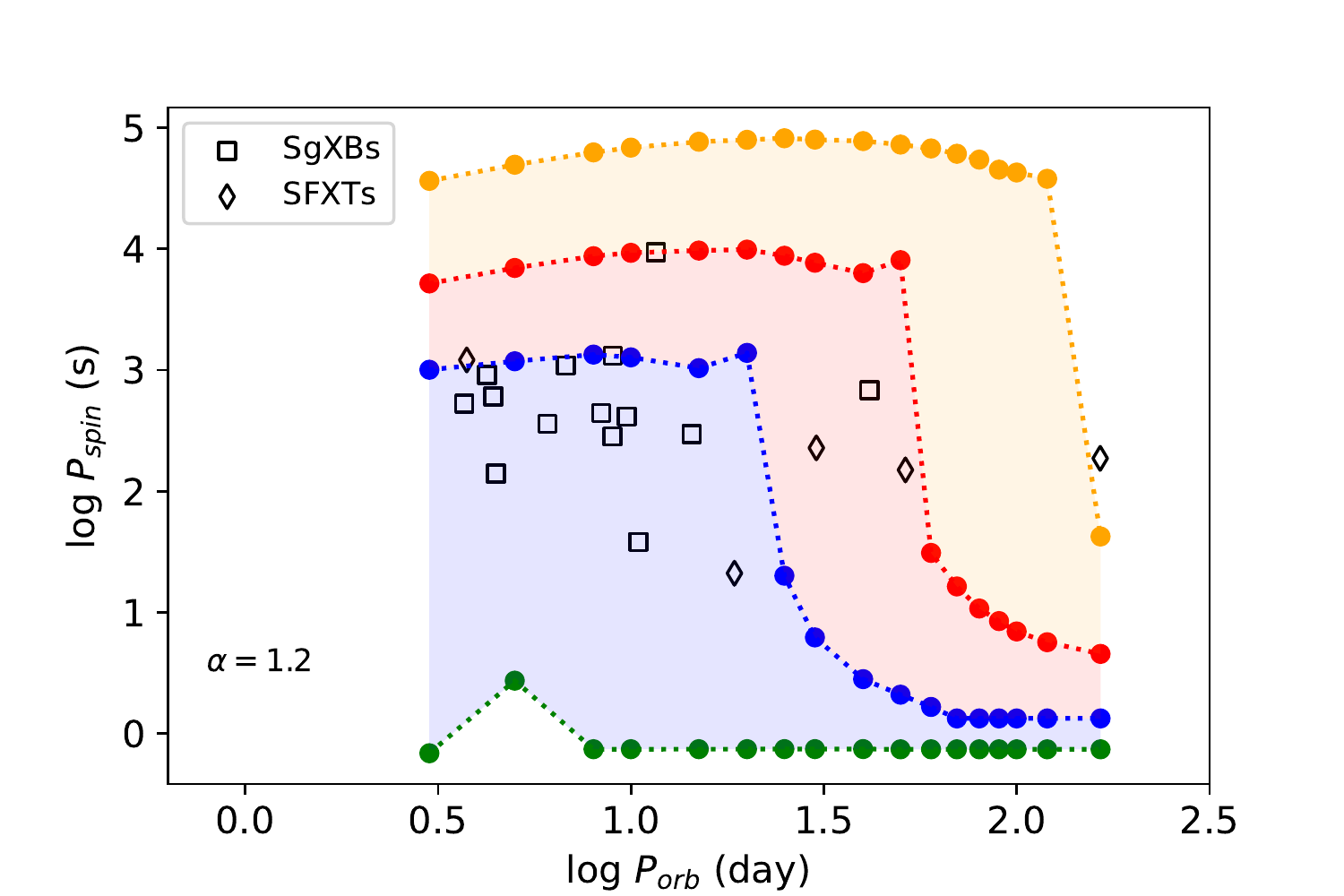}
	\includegraphics[width=0.5\textwidth]{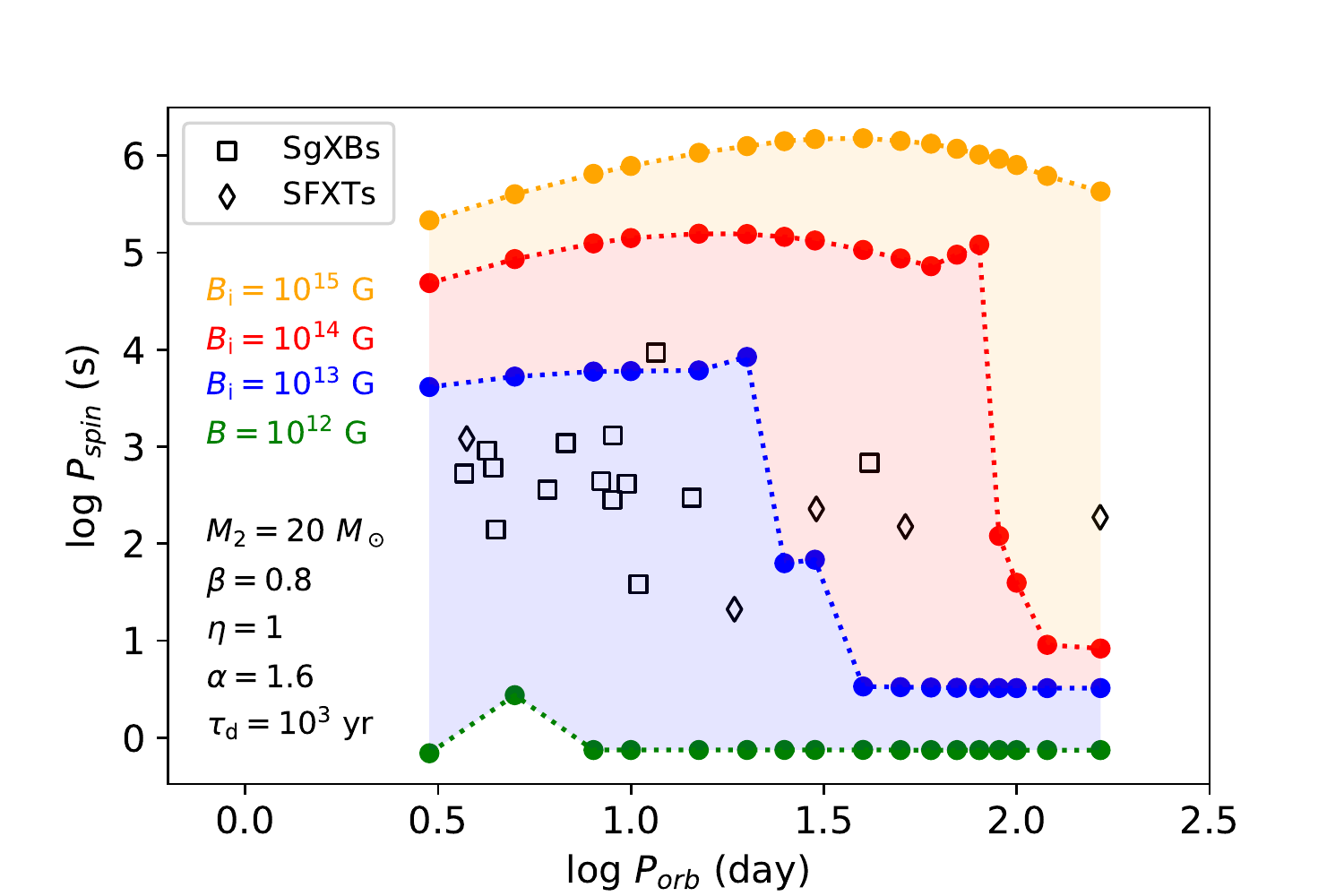}
	\includegraphics[width=0.5\textwidth]{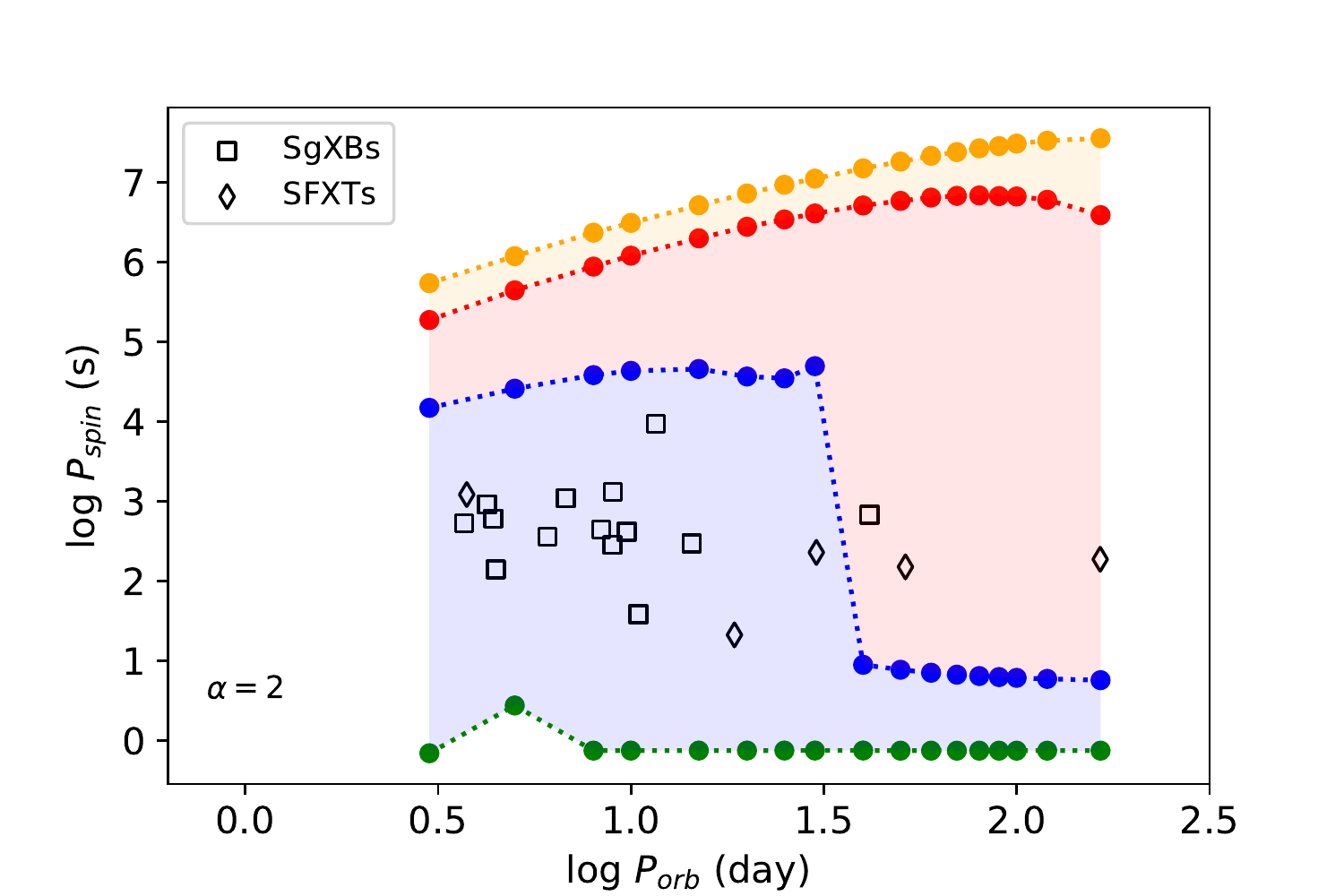}
	\caption{Same as Figure \ref{fig:m20_bt} but with $\alpha=$1, 1.2, 1.6 (the reference model) and 2 from top-left to bottom-right.}
	\label{fig:alpha_tau3}
\end{figure}

\begin{figure}
	\includegraphics[width=0.5\textwidth]{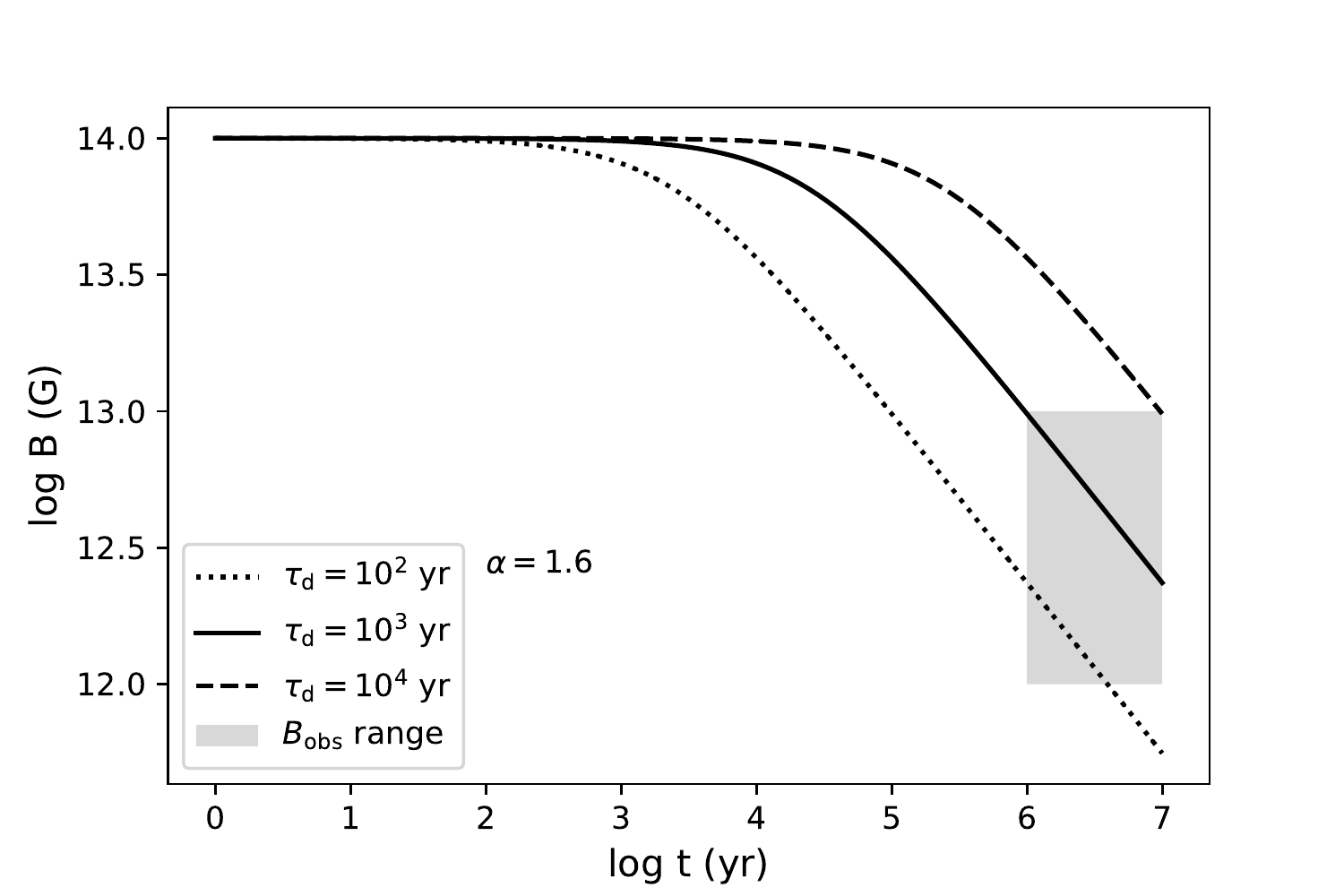}
	\includegraphics[width=0.5\textwidth]{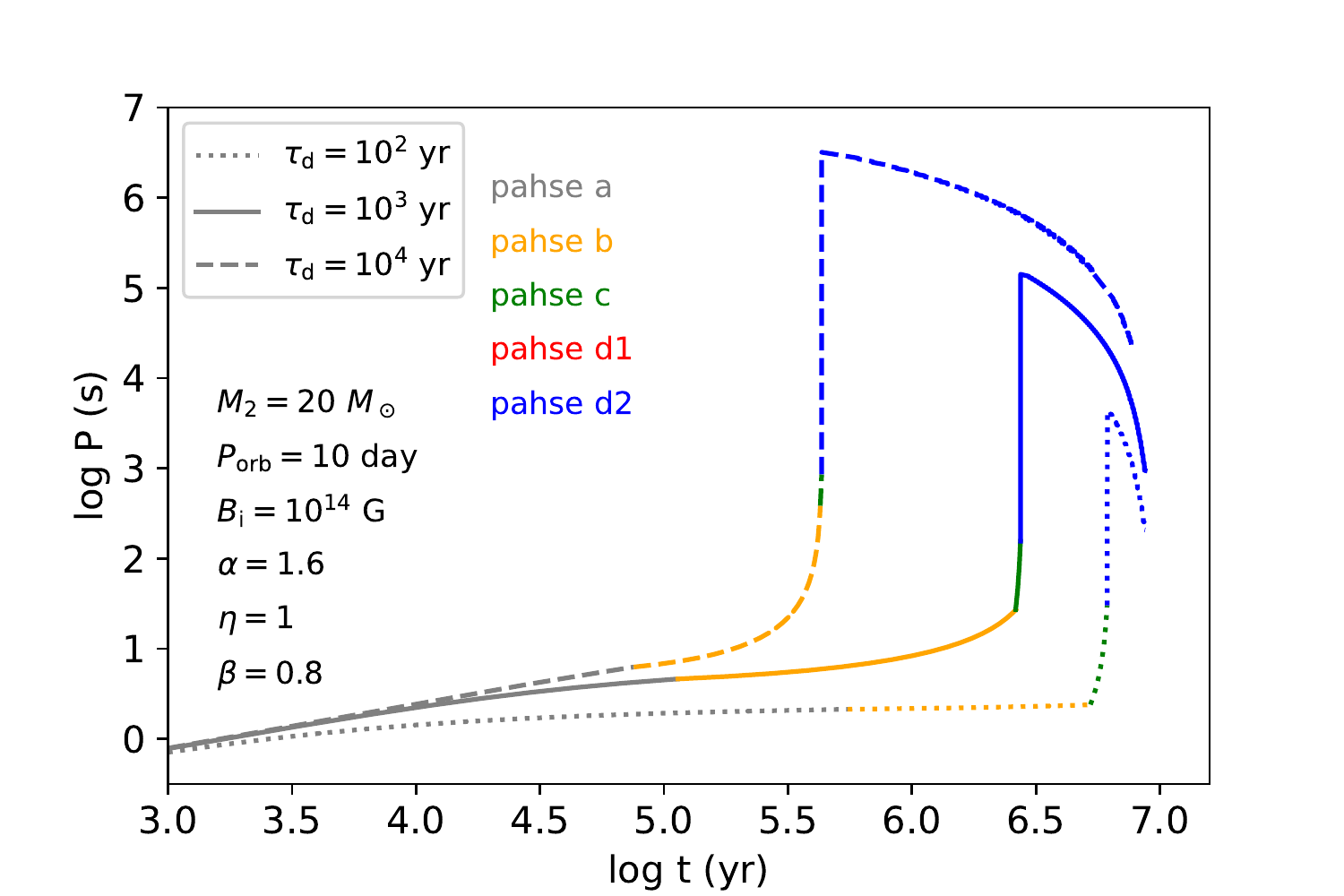}
	\caption{The magnetic field (left panel) and spin (right panel) evolution tracks with diffirent $\tau_{\rm d}$. The gray region in the left panel covers the range of the estimated magnetic field and age of NSs in HMXBs from observations.}
	\label{fig:BP_evo_tau}
\end{figure}

\begin{figure}
	\centering
	\includegraphics[width=0.5\textwidth]{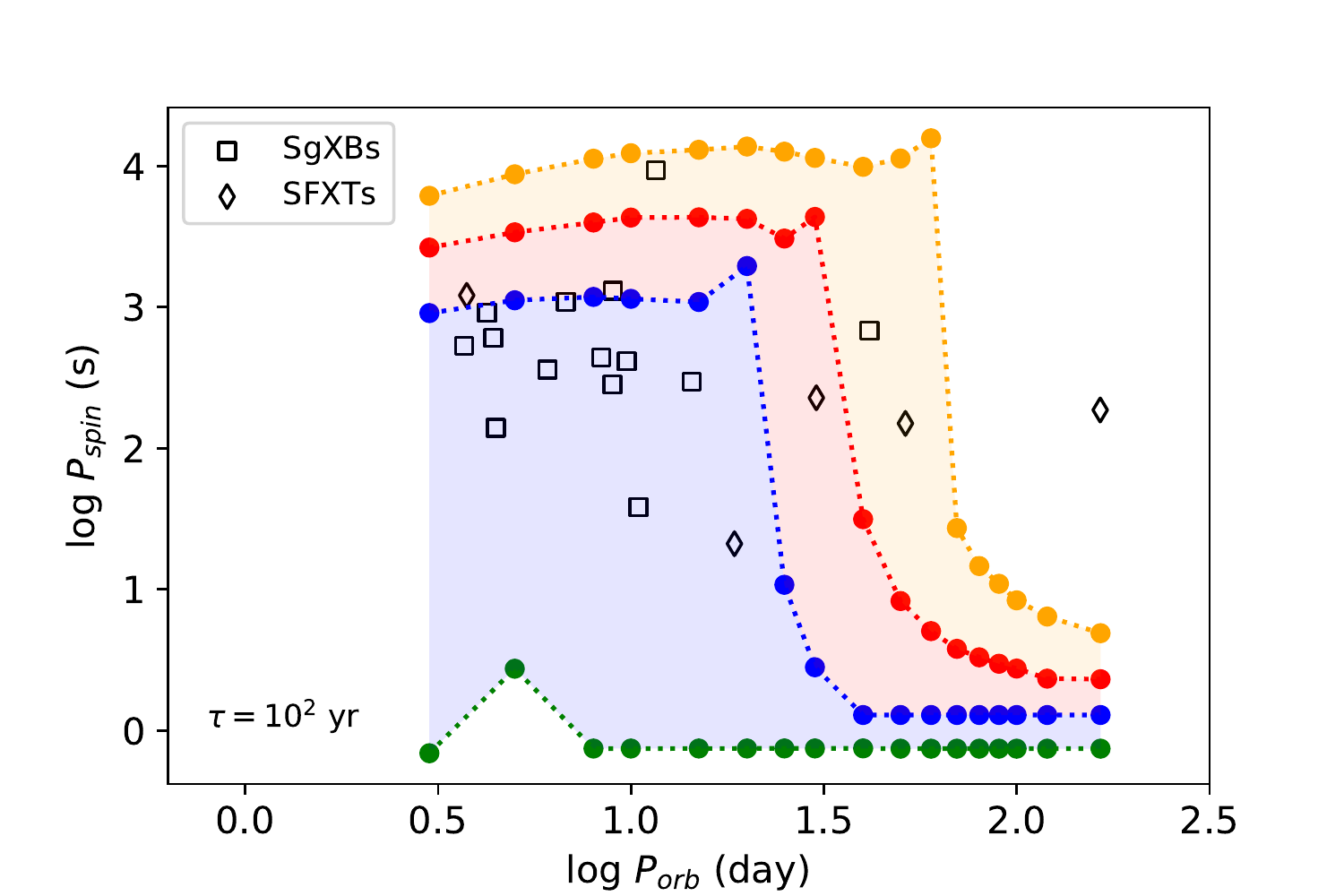}
	\includegraphics[width=0.5\textwidth]{a16_tau3.pdf}
	\includegraphics[width=0.5\textwidth]{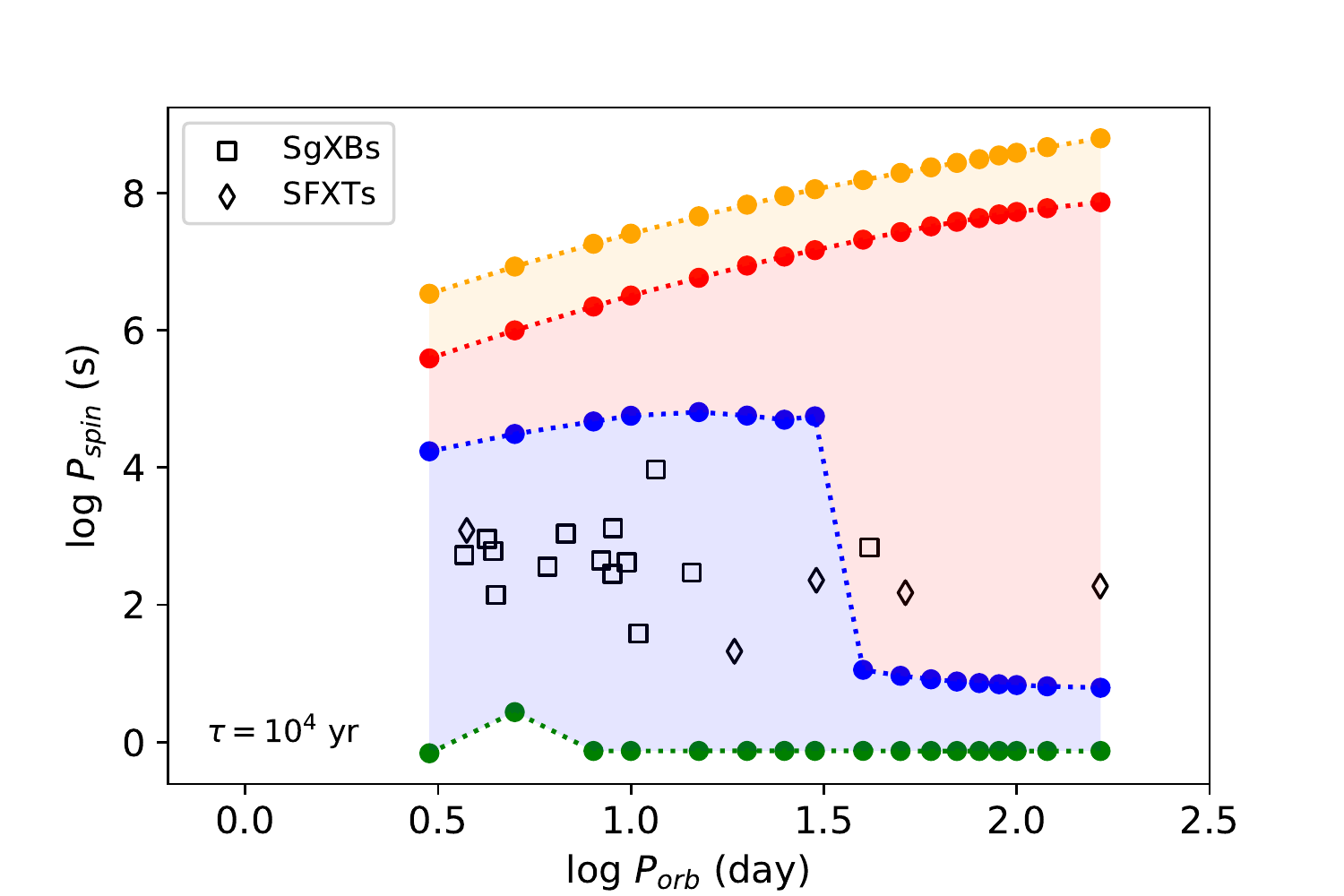}
	\caption{Same as Figure \ref{fig:m20_bt} but with $\tau_{\rm d}=10^2$, $10^3$ (the reference model) and $10^4$ yr, from top to bottom.}
	\label{fig:alpha16_tau}
\end{figure}

We take $\alpha=1.6$ and $\tau_{\rm d,i}=\tau_{\rm d}/(B_{\rm i}/10^{15} {\rm ~G})^{\alpha}$ in the reference model, where $\tau_{\rm d}=10^{3} {\rm ~yr}$. 
There are some debates on the range of $\alpha$, e.g., \cite{DallOsso2012} takes $0 \leq \alpha \leq 2$ while \cite{Fu2012} thinks $1.5 \lesssim \alpha \lesssim 1.8$ is most likely.
We plot the magnetic field and spin evolution tracks with different $\alpha$ and $\tau_{\rm d}$ in Figure \ref{fig:BP_evo_alpha} and \ref{fig:BP_evo_tau}, where the gray region covers the range of the  estimated magnetic fields \citep{Caballero2012} and ages of NSs in HMXBs from observations.
They show in the cases with large $\alpha$ and $\tau_{\rm d}$, which means the magnetic fields evolve slowly, the NSs can spin down to long periods.
The results of the longest spin periods with different $\alpha$ and $\tau_{\rm d}$ are exhibited in Figure \ref{fig:alpha_tau3} and \ref{fig:alpha16_tau}, respectively.
They show that more NSs need strong magnetic fields to explain their long spin periods in the rapid evolution cases (with small $\alpha$ and $\tau_{\rm d}$), while in the slowly evolution cases, magnetar model is dispensable for the observed sources.

\bibliographystyle{raa}
\bibliography{pp_hmxb_ref}

\end{document}